\documentclass{article}

\usepackage{arxiv}
\usepackage[utf8]{inputenc}
\usepackage[T1]{fontenc}
\usepackage{graphicx}
\usepackage{amsmath}
\usepackage{booktabs}
\usepackage{multirow}
\usepackage[authoryear,round]{natbib}
\usepackage{caption}
\usepackage{array}
\usepackage{longtable}
\usepackage{placeins}
\usepackage{hyperref}

\graphicspath{{figures/}}

\renewcommand{\headeright}{Terwilliger et al.}
\renewcommand{\undertitle}{A Preprint}
\renewcommand{\shorttitle}{Invisible walls}

\newcommand\blfootnote[1]{%
  \begingroup
  \renewcommand\thefootnote{}%
  \footnotetext{#1}%
  \endgroup
}

\title{Invisible walls: how pedestrians navigate around social interactions}

\author{
  \mdseries
  {\normalsize Jack Terwilliger$^{1*}$,\; Julia Di Silvestri$^{2}$,\; Seika Murase$^{1}$} \\
  {\normalsize Anne Elizabeth Clark White$^{3}$,\; Federico Rossano$^{1}$} \\[0.4em]
  {\small $^1$Dept.\ of Cognitive Science, UC San Diego \quad $^2$Dept.\ of Biology, UC San Diego} \\
  {\small $^3$Herbert Wertheim School of Public Health \& Human Longevity Science,} \\
  {\small UC San Diego}
}

\date{}

\begin{document}

\maketitle

\blfootnote{$^{*}$Corresponding author: Jack Terwilliger, \href{mailto:jterwilliger@ucsd.edu}{\texttt{jterwilliger@ucsd.edu}}.}

\begin{abstract}
Pedestrian dynamics are characterized as complex physical systems constrained by social norms, like personal space. Yet, prior research has ignored the spatial norms imposed by others' social interactions, such as conversations. Unlike physical obstacles, the boundaries of social interactions are latent --- even more so than those of personal space --- as if they are invisible walls, and must be inferred from signs of interactional involvement. In four field experiments with 4,911 participants, we show that pedestrians integrate others' gaze, proximity, body orientation, and talk to avoid interrupting possible interactions. However, pedestrians also collectively violate these spatial norms by walking through an interaction if other pedestrians had already done so. These results demonstrate how physical mobility depends on pedestrians' social computations.
\end{abstract}


\section*{Introduction}

Pedestrian traffic is an integral part of daily life, situated at the intersection of human behavior, the built environment, and, increasingly, autonomous vehicles and robots. Understanding the dynamics of pedestrian traffic is essential for a variety of human-centered design problems, but it is a complex phenomenon to model. As such, it has been the topic of considerable multidisciplinary research \citep{1} spanning physics \citep{2,3,4}, physiology \citep{5}, engineering \citep{6}, and the social sciences \citep{7,8}.

To predict pedestrian traffic, researchers often characterize pedestrians as self-propelled elements of a complex physical system. Individually, pedestrians take energetically efficient trajectories toward their goals \citep{5,9} and anticipate the trajectories of other pedestrians and objects to avoid collisions \citep{10,11}. Collectively, pedestrians form lanes \citep{2}, synchronize their movement \citep{6,12}, and oscillate as a crowd \citep{3}. Often, these dynamics are modeled as systems of particles \citep{4}, fluids \citep{3}, or plasmas \citep{2}.

\begin{figure}[t!]
  \centering
  \includegraphics[width=\linewidth]{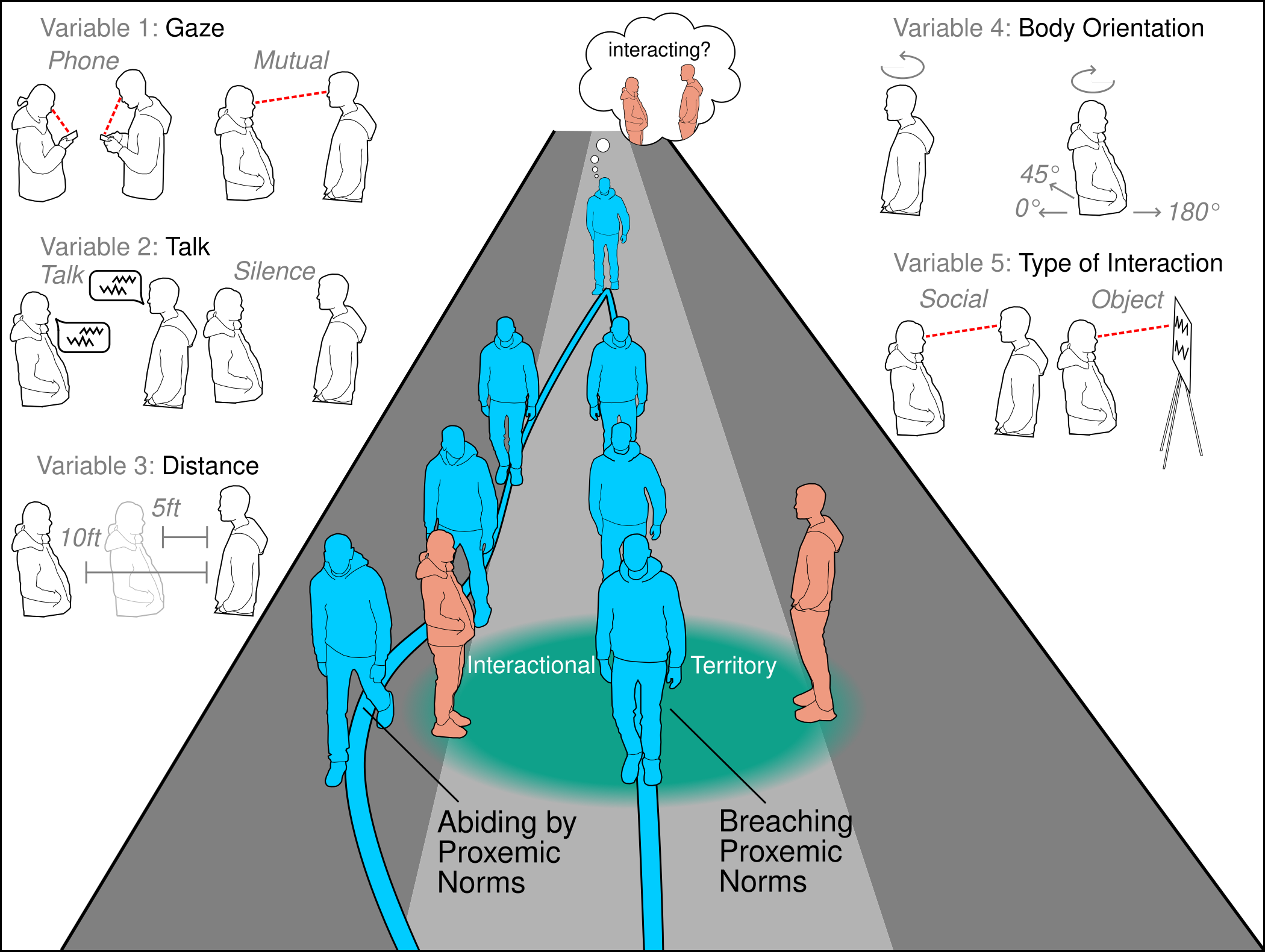}
  \caption{A diagram of our field experiments. Actors (shown in orange) stood in busy pathways while displaying 5 distinct signs of interactional involvement (see depictions of Variables 1--5). In different configurations of these variables, we observed whether pedestrians (shown in blue) walked between or around our actors and we measured pedestrians' walking trajectories.}
  \label{fig:1}
\end{figure}

But, people are not particles: they are social animals. Biological research has long revealed that animals' spatial navigation skills and behavior are shaped by the demands of their socio-ecology \citep{13,14,15,16,17}. For humans, one of the key demands of spatial navigation is the frequent need to traverse public places and therefore coming into contact with other humans. As such, an understanding of pedestrian behavior needs to take into account a broad range of human social dynamics.

On the other hand, the social sciences characterize pedestrian traffic as tracing the contours of human sociality. For example, walking together is an emblematic joint commitment underpinned by interpersonal rights and obligations \citep{7,8}. But crucially, pedestrians also comply with proxemic norms: tacit rules that regulate the social use of space \citep{18}. Research has focused on how pedestrians avoid intruding into each other's personal space \citep{19,20}, defined as ``an area with invisible boundaries surrounding a person's body into which intruders may not come'' \citep{21}. The extent of personal space is culturally determined \citep{18} and shaped, in each social encounter, by relations of race \citep{22,23} and gender \citep{23,24,25}.

Yet, research on pedestrian social dynamics has mostly focused on properties of the individuals pedestrians encounter (e.g., their kinematics, ethnicity, gender), while largely neglecting the activities those individuals are engaged in, especially their social interactions, such as conversation. Consider, for example, two individuals chatting in front of a busy library entrance; even if the space between them is physically unoccupied or claimed by personal space, pedestrians will tend to walk around the interactants rather than between them (Fig.\ 1, Movie~S1\footnote{Movie~S1: \url{https://jackft.github.io/iw/supplementary/SM1_demonstration.mp4}}). While prior work has characterized the spatial arrangements interactants adopt during face-to-face encounters, such as F-formations \citep{26}, it has not investigated how pedestrians navigate around social interactions in naturalistic settings. Such activities impose an interactional territory \citep{27}: an invisible boundary around an interaction into which non-participants may not come.

In order for pedestrians to navigate around social interactions, they must recognize the presence, extent, and claims of interactional territory. However, unlike physical obstacles and to a greater degree than personal space, interactional territories are not directly observable. We propose that pedestrians infer the presence and extent of interactional territory by reading others' involvement signs \citep{28}. These include where others are looking, whether they are talking, how their bodies are oriented, how far apart they are standing, where they are standing, and so on. Conversation is a particularly relevant case for pedestrians: humans spend roughly 30\% of their waking hours in conversation \citep{29}, and conversation has broadly shaped human cognition \citep{30}. Consequently, our perceptual systems are especially vigilant at detecting conversational interaction \citep{33,34,35,36}. At the same time detecting involvement is, evidently, a challenging problem to for machines \citep{31,32}.

Furthermore, we propose that once pedestrians detect interactional territory, they must then reason about the social and energetic costs of possible walking trajectories. On one hand, politeness motivates pedestrians to avoid breaching others' interactional territory, since they would impede and interrupt those people. Therefore, doing so constitutes a negative face-threatening act \citep{37,38}. On the other hand, the least socially costly trajectory may not be the least energetically costly trajectory, and so pedestrians are motivated to breach interactional territory. The collective behavior of pedestrians may facilitate the breaching of interactional territory. Decisions to breach social norms are the product of both individual and collective decision-making \citep{39}. Pedestrians' violations of explicit laws such as jaywalking \citep{40} and trespassing \citep{41} are shaped by the behavior of other pedestrians. Whether a pedestrian breaches an interactional territory, then, may depend not only on how much involvement the interactants display, but also on whether preceding pedestrians have already breached it.

Here, we investigate how pedestrians navigate around others' social interactions using 4 breaching experiments \citep{42}. In each experiment, we positioned a pair of actors on heavily trafficked public pathways where they were instructed to display their involvement in interaction with each other or with objects. We then observed how many pedestrians walked between or around the actors. In 2 of these experiments, we also measured pedestrian trajectories as they navigate in these spaces.

We show that pedestrians are acutely aware of others' social interactions and the proxemic norms they impose. The probability pedestrians breach these norms depends on how others display their involvement in social interaction, specifically, their body orientation, interpersonal distance (5ft vs. 10ft), whether they shared mutual gaze, and whether they were speaking. These effects cannot be explained by gaze aversion or physical occupancy. Finally, we show that pedestrians engage in collective rule-breaking, where pedestrians are more likely to breach a proxemic norm when a preceding pedestrian has done so.

\section*{Materials and Methods}

This study was approved by the University of California, San Diego Institutional Review Board (161452). Participants in experiment 4 gave informed consent. For experiments 1--3, consent was waived since observations were made at heavily trafficked public campus locations where surveillance cameras were already present and visible, and obtaining consent would have required disruptively stopping pedestrians. Full procedures are given in the Supplementary Materials.

\textbf{Experiments 1 and 3.} We positioned actors on heavily trafficked campus pathways and observed how many pedestrians passed between (breached) versus around them. In experiment 1, two actors stood 5 or 10 feet apart in a $2 \times 2 \times 2$ design manipulating mutual gaze (vs.\ looking at a phone), talk (vs.\ silence), and interpersonal distance, while also varying actor gender; we filmed 2,319 pedestrians (2017--2019). In experiment 3, one actor stood in front of an informational sign or an art mural and either gazed at it or down at a phone; we observed 1,515 pedestrians (2018--2019). Actors switched conditions on a counterbalanced schedule. Trained research assistants manually coded pedestrian attributes (presumed gender, mode of transit, group membership, distraction) and breaching outcomes from video (coding schemes in Tables~S1--S2); coders achieved 89.8\% mean accuracy against a gold-standard set.

\textbf{Experiment 2.} Two actors stood 10 feet apart on a busy pathway while we manipulated their mutual body orientation (face to face, back to back, 45\textdegree{} offset, or a no-actor baseline); we filmed 978 pedestrians (2023) using two pole-mounted cameras composited into a single synchronized video. In addition to the coding above, we measured continuous pedestrian trajectories (see below) and, for the analysis of collective breaching, counted local crowding and the number of pedestrians who had breached in the 5\,s before each pedestrian passed.

\textbf{Experiment 4.} We recruited 99 undergraduates (46 men) under the cover story of a visual-spatial memory task where they had to look at an outdoor mural and redraw it from memory when they returned to the lab (2022). On their return through an outdoor hallway, we placed either two women actors, two chairs, or nothing (baseline) across the entrance and recorded whether participants passed between or around them, tracking their trajectories to measure how close they approached before diverting.

\textbf{Statistical analysis.} For each observational experiment we fit generalized linear mixed models (GLMMs) with a binomial link predicting breach (0/1) from the manipulated involvement signs, with random intercepts for location and date (and minute-within-hour for experiment 2), using the \textit{lme4} R package \citep{51}. Effects are reported as odds ratios from estimated marginal means computed with \textit{emmeans} \citep{52}. Experiment 4 was fit with a Bayesian GLMM using \textit{brms} \citep{57} because of complete separation in the baseline and chair conditions. All analyses used R 4.5.1. Full model specifications and diagnostics are given in the Supplementary Materials.

\textbf{Trajectory measurement.} We recovered continuous world-coordinate trajectories via a computer-vision pipeline (Fig.\ S5). We detected pedestrians and their feet with YOLOv7 \citep{55}, tracked them across frames with SORT \citep{56} followed by manual correction, and imputed gaps with a Kalman filter and Rauch--Tung--Striebel smoothing. A homography mapping image pixels to a ground plane converted foot positions to world coordinates (mean hold-one-out reprojection error 15.48\,cm).

\textbf{Data and code availability.} Data, analysis code, and supplementary videos are available at \url{https://github.com/jackft/iw}.

\section*{Results}

\subsection*{Displays of social involvement differentially redirect pedestrian flow}

Face-to-face conversation is a cornerstone of society, and its conduct depends on speech, interpersonal distance \citep{18}, and gaze \citep{43}. Here, we investigate how these signs of others' involvement in social interaction affect pedestrian traffic.

In experiment 1, we placed two actors on 3 pathways at a university campus. We used a $2 \times 2 \times 2$ design in which we manipulated whether the actors were looking at each other vs.\ looking at a phone, whether they were talking vs.\ silent, and whether they stood 5 vs.\ 10 feet apart (see Fig.\ 2A); we also varied actor gender (both men vs.\ both women vs.\ one man and one woman). We manually coded pedestrian gender, mode of transportation (e.g.\ walking vs.\ scootering), possible distractions (looking at a smartphone or talking to someone), and whether pedestrians appeared to be traveling alone or together. We filmed 2,319 pedestrians (1,335 men, 921 women, 63 unsure) between April 2017 and January 2019, excluding those whose gender was `unsure' or who never traversed near part of the path with our actors. These exclusions resulted in 1,861 pedestrians (1,110 men).

\begin{figure}[tbp]
  \centering
  \includegraphics[width=\linewidth]{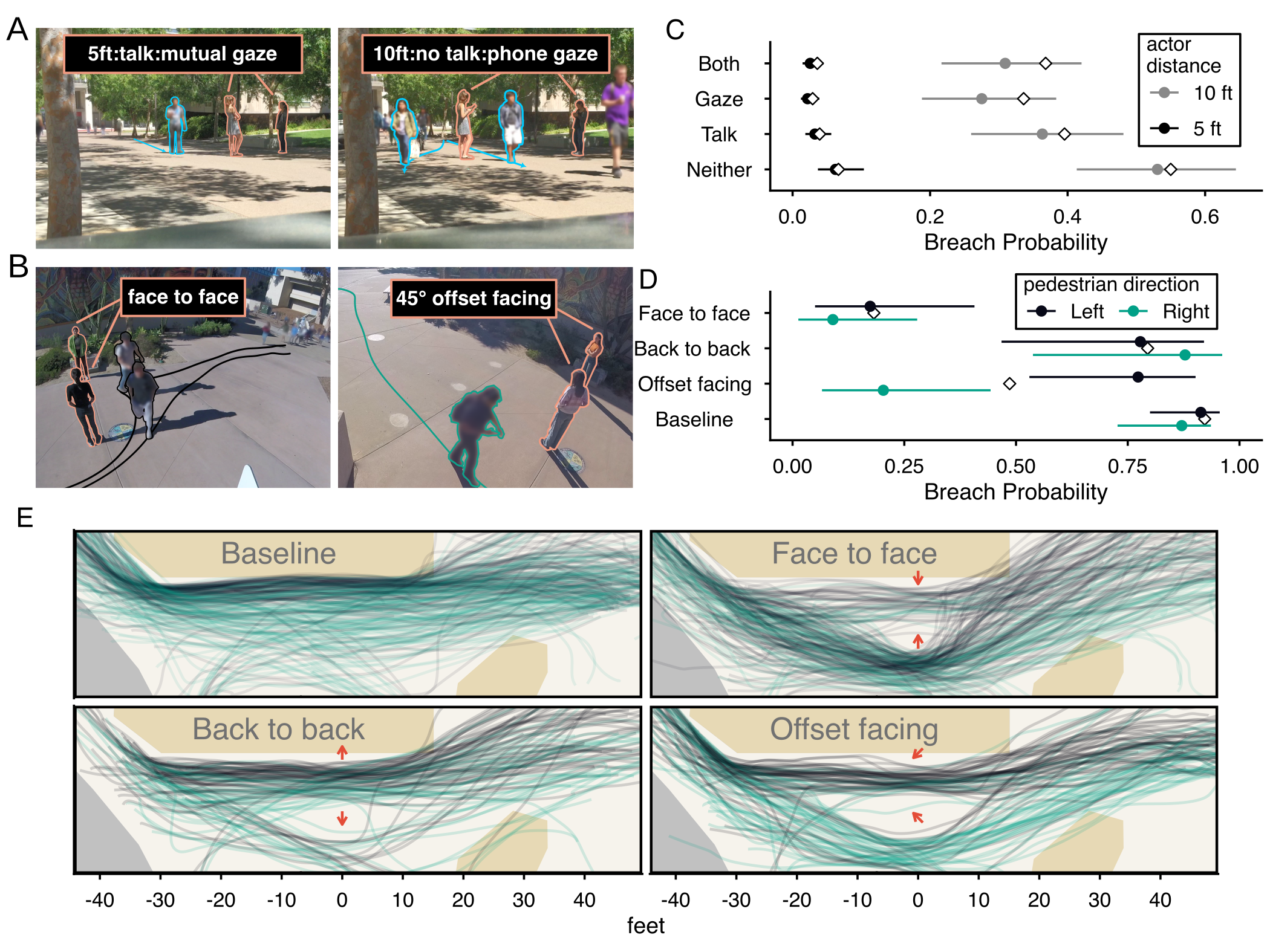}
  \caption{(\textbf{A}) Video frames from experiment 1.
  (\textbf{B}) Video frames from experiment 2.
  (\textbf{C}) Breaching probabilities from experiment 1. Dots represent the estimated marginal mean and lines represent 95\% confidence intervals. Empirical breaching rates are represented with white diamonds.
  (\textbf{D}) Breaching rates from experiment 2 plotted by condition and pedestrian walking direction. Conditions are depicted on the right side of the plot.
  (\textbf{E}) Pedestrian trajectories from experiment 2 are plotted and colored by direction. Green is rightward. Black is leftward. Actor body orientation is depicted by red arrows. Types of terrain are represented by the colored background. Off-white depicts a pedestrian pathway, yellow depicts patches of dirt and vegetation, and grey depicts an asphalt roadway.}
  \label{fig:2}
\end{figure}

Pedestrians were 1/3 as likely to breach when the actors shared mutual gaze than when the actors looked down at their phones (OR = 0.34 $\pm$ 0.07, $z = -5.58$, $P < 0.001$). They were 1/2 as likely to breach when the actors talked than when the actors were silent (OR = 0.51 $\pm$ 0.10, $z = -3.76$, $P = 0.001$), and were 1/17 as likely to breach when the actors halved their interpersonal distance, i.e., when they stood 5 feet apart than when they stood 10 feet apart (OR = 0.06 $\pm$ 0.01, $z = -12.21$, $P < 0.001$) (see Fig.\ 2C). Notably, we observed a subadditive interaction between actor mutual gaze and talk (OR = 2.34 $\pm$ 0.64, $z = 3.10$, $P = 0.002$): pedestrians were no less likely to breach when the actors shared mutual gaze and talked than when they shared mutual gaze in silence (OR = 1.18 $\pm$ 0.24, $z = 0.81$, $P = 0.419$). This pattern suggests that, rather than integrating talk and gaze additively \citep{44}, pedestrians' belief about actors' involvement was saturated by mutual gaze. Together, these results suggest that pedestrians integrate information across modalities -- gaze, speech, and proximity -- to infer the type and level of involvement. Merely tracking others' mutual body orientation is not sufficient, since this was held constant across all conditions. Body orientation, however, is an important resource for initiating \citep{45}, maintaining \citep{26}, and conducting conversations \citep{18,43} and is made highly salient by the human visual system \citep{33,46}.

In experiment 2, we manipulated the actors' mutual body orientation by placing two actors on a busy pathway on a university campus. The actors stood directly face to face, back to back, or so that they were both 45 degrees offset from the face to face condition (see Fig.\ 2B\&E). We also observed pedestrian traffic with no actors present. We coded pedestrian observations similar to experiment 1 but also tracked their trajectories (see Methods). We filmed a total of 978 pedestrians (458 men) at various times of day between November and December in 2023. We excluded pedestrians who did not cross the space where the actors stood, yielding 686 pedestrians (301 men).

When the actors stood face to face, pedestrians were 1/6 as likely to breach compared to when the actors stood 45 degrees offset (OR = 0.17 $\pm$ 0.08, $z = -3.61$, $P < 0.001$), 1/34 as likely compared to when the actors stood back to back (OR = 0.03 $\pm$ 0.02, $z = -4.84$, $P < 0.001$), and 1/62 compared to the baseline condition (OR = 0.016 $\pm$ 0.01, $z = -7.23$, $P < 0.001$). We also observed a significant interaction between pedestrian walking direction and actor body orientation when they stood 45 degrees offset: pedestrians were 1/13 as likely to breach when they approached our actors head-on vs.\ from behind (OR = 0.08 $\pm$ 0.04, $z = -4.38$, $P < 0.001$) (see Fig.\ 2D). Finally, we observed that pedestrians who were using their phones while approaching the actors were twice as likely to breach compared to those who were not using their phones (OR = 2.01 $\pm$ 0.69, $z = 2.03$, $P = 0.042$), meaning that pedestrians' own body orientations were also integrated into navigation decisions. Together, these results suggest that pedestrians actively monitor their environment, using others' mutual body orientation, to infer the presence of others' social interactions.

\subsection*{Pedestrian flow depends on with what and with whom others are interacting}

We proposed that pedestrians are unlikely to walk between people sharing mutual gaze and directly facing each other because pedestrians make inferences about social interactions and the proxemic norms they impose. An alternative account is that pedestrians are only making inferences about how these displays relate to pedestrians themselves. For example, our results could be the result of gaze avoidance, an evolutionarily conserved sensitivity many species exhibit to being gazed at or being in the direction someone is facing \citep{47}.

We investigated this in experiment 3. We placed one actor in front of either an informational sign or an art mural and manipulated whether the actor was looking up/out at the sign versus down at their phone. We also varied the gender of the actor (man vs.\ woman). Experiment 3 was conducted at the same 4 locations as experiments 1 and 2. We filmed 1,515 pedestrians (889 men, 593 women, 33 unsure) between August 2018 and April 2019, excluding those whose gender was `unsure' or who never traversed near part of the path with our actors. These exclusions resulted in 874 pedestrians (528 men).

\begin{figure}[tbp]
  \centering
  \includegraphics[width=0.5\linewidth]{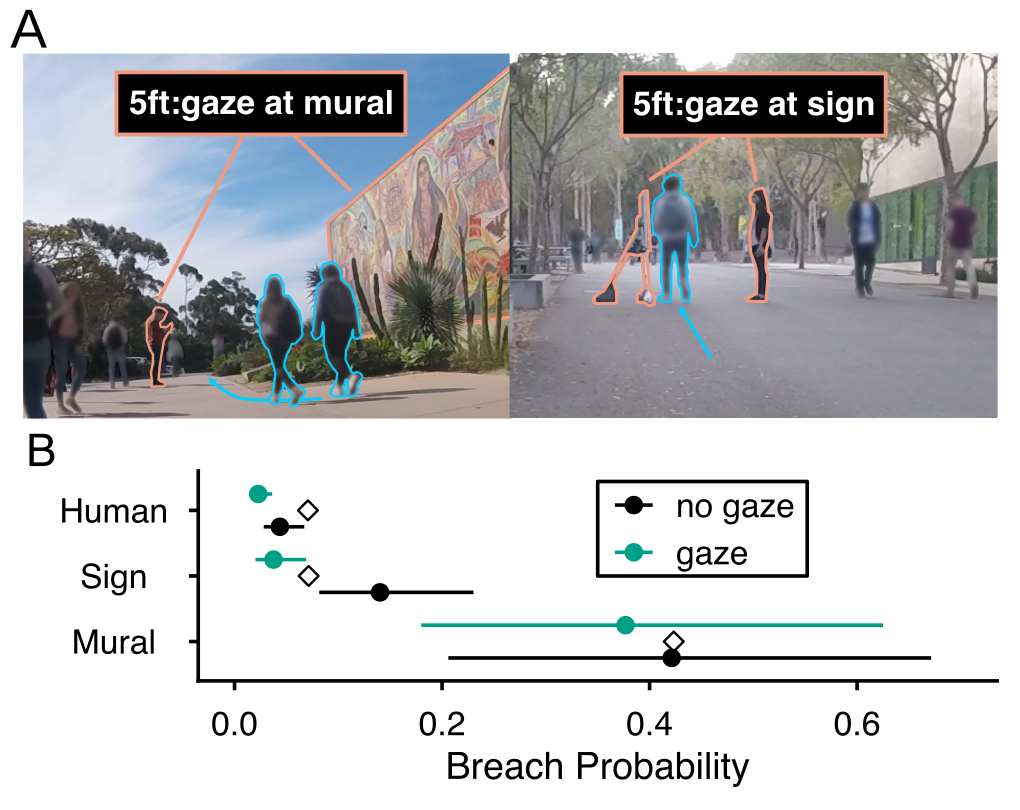}
  \caption{(\textbf{A}) Video frames from experiment 3 of the 5\,ft + mutual gaze + sign (left) and the 5\,ft + no gaze + sign condition (right).
  (\textbf{B}) Breaching probabilities pooled from experiment 1 \& 3 at 5 feet. Dots represent the estimated marginal mean and lines represent 95\% confidence intervals. Empirical breaching rates are represented with white diamonds.}
  \label{fig:3}
\end{figure}

Crucially, after pooling data from experiments 1 and 3 (see Fig.\ 3), we observed an additive interaction between the sign condition and directed gaze (OR = 0.47 $\pm$ 0.12, $z = -3.03$, $P = 0.003$). While pedestrians were 3.57 times more likely to walk between an information sign and an actor than between two actors when actors looked down at their phones (OR = 3.57 $\pm$ 1.20, $z = 3.80$, $P < 0.001$), pedestrians were no more likely to breach in the sign condition than the human condition when actors directed their gaze at a sign/each other (OR = 1.67 $\pm$ 0.56, $z = 1.52$, $P = 0.128$). Together, these results show that pedestrians do not just take the direction of others' gaze into account when navigating. They make inferences about what others are looking at and what type of interaction they are involved in.

We investigated this in experiment 4. We placed either two actors or two chairs facing each other, in front of an outdoor hallway, 5 feet apart (a distance which approximately obstructed the hallway) (see Fig.\ 4); additionally, we ran a baseline condition, with no objects or actors. Unlike our other field experiments, we recruited 99 undergraduate participants (46 men). We obfuscated the aim of the study by asking them to walk outside from a testing room to an outdoor mural and draw it from memory after they returned. Pedestrians were substantially less likely to walk between the actors than they were between two chairs (OR = $1.05 \times 10^{3}$, 95\% CI = [$1.67 \times 10^{1}$, $1.82 \times 10^{4}$]). 100\% of participants in the baseline or chairs conditions walked between whereas only 11.7\% ($N = 6$) pedestrians did so in the actors condition ($N_{\text{baseline}} = 26$, $N_{\text{chairs}} = 22$, $N_{\text{actors}} = 51$).

We tested the effects of gender on proxemic breaching. In experiment 1, pedestrians were twice as likely to breach the interactional territory of women than men (OR = 2.12 $\pm$ 0.61, $z = 2.60$, $P = 0.009$), but, in experiment 3, there was no gender difference (OR = 1.01 $\pm$ 0.20, $z = 0.06$, $P = 0.953$). We found no effects of pedestrian gender in any of experiments 1--4. In experiment 4, we asked whether pedestrian men were more likely to expect actor women to yield their interactional space or suspend their interactions than pedestrian women. We tested whether men were closer to the actors when their trajectories changed course to walk around the actors (see Fig.\ 4). We found no evidence of this ($\beta_{\text{man}} = -2.45 \pm 4.12$ feet, $t = -0.60$, $P = 0.557$). See supplementary material for further context on these gender effects.

\begin{figure}[tbp]
  \centering
  \includegraphics[width=0.5\linewidth]{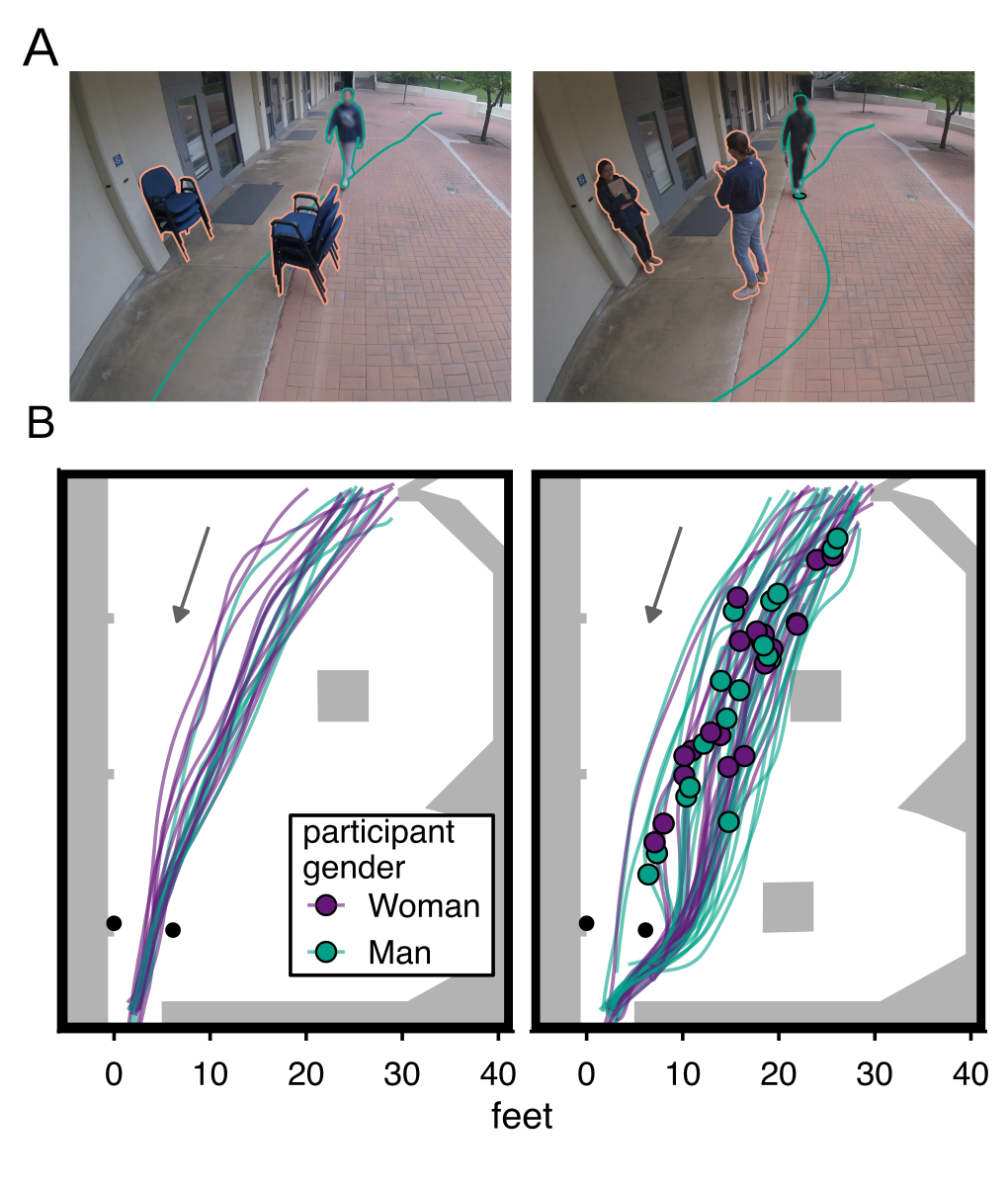}
  \caption{(\textbf{A}) Video frames from experiment 4.
  (\textbf{B}) Pedestrian trajectories from experiment 4 colored by pedestrian gender (green for man, purple for woman). Colored circles mark where each participant changed course around the actors.}
  \label{fig:4}
\end{figure}

\subsection*{Pedestrians collectively breach interactional territory}

We further tested whether pedestrians collectively breach proxemic norms as a result of social contagion -- a phenomenon where a behavior spreads to individuals as a consequence of exposure to others' enacting that behavior. We controlled for the physical effects of crowding and for temporal effects of class schedules which might correlate pedestrians' behavior without constituting social contagion. For example, with substantial crowding (a) our actors' interactional territory may be the only free space to pass through or (b) pedestrians may, independently, infer that our actors have weak claims to their interactional territory because they are impeding transit of the many.

We investigated social contagion by analyzing the data in experiment 2, in which we tracked 686 pedestrian trajectories. For every pedestrian, we counted the number of other pedestrians who breached 5 seconds prior to when they passed the actors. We counted the number of pedestrians present at the moment they passed the actors. Finally, we included random intercepts for the date and which 10-minute interval within the hour our observations were made.

We observed that pedestrians were over twice as likely to breach if a preceding pedestrian had also breached 5 seconds prior compared to when no preceding pedestrians had done so (OR = 2.77 $\pm$ 0.58, $z = 4.88$, $P < 0.001$). Previous research has shown a similar collective rule-breaking effect about jaywalking, e.g., a pedestrian is 1.5--2.5 times more likely to jaywalk when a preceding pedestrian has \citep{40}. See Fig.\ 5 for a qualitative example of collective proxemic rule-breaking.

\begin{figure}[tbp]
  \centering
  \includegraphics[width=\linewidth]{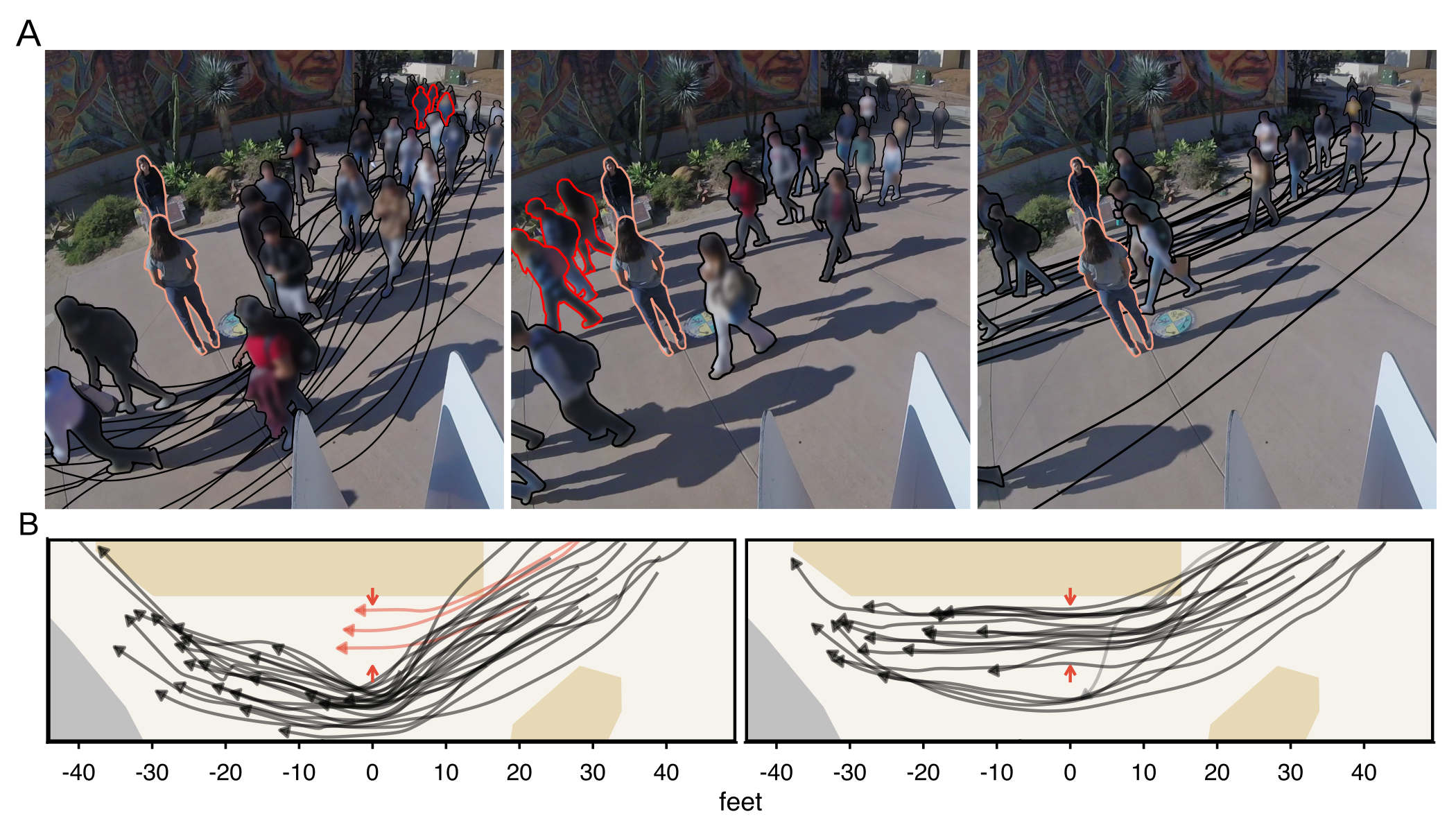}
  \caption{On December 7th, 2023 at 12:31:35, over the span of 35.5 seconds, a crowd of 46 pedestrians walked past an art mural while two actors silently stood 10 feet apart while facing and staring at each other. The first 22 pedestrians walk around the actors. 13.6 seconds after the arrival of the first pedestrian, a group of 3 pedestrians (in red) walk between the actors within 0.1 seconds of each other. Just 2.1 seconds later, the first of 12 of the 19 remaining pedestrians also walk between them.
  (\textbf{A}) Video frames before, during, and after the group of 3 pedestrians breaches.
  (\textbf{B}) Pedestrian trajectories before and after the group of 3 pedestrians breaches.}
  \label{fig:5}
\end{figure}

\FloatBarrier
\section*{Discussion}

Robert Sommer, who pioneered the study of personal space, wrote that ``the best way to learn the location of invisible boundaries is to keep walking until somebody complains'' \citep{21}. We inverted this methodology to investigate interactional territory. We showed that pedestrians make fine-grained inferences about others' social interactions. The probability that pedestrians breach an interactional territory depends systematically on both how others display their involvement in these activities and whether other pedestrians have already breached these territories. Unlike physical collisions, breaching proxemic norms does not just perturb the order of traffic --- it also perturbs the social order. We argue that pedestrian traffic reflects the rich human social umwelt \citep{48} --- our species specific way of experiencing the world --- as much as it reflects the physical dynamics of complex systems composed of people. This has consequences for human-centered design. Autonomous robots are increasingly designed to share pedestrian environments \citep{20,49,50}, and therefore they must be able to model the human social umwelt if they are to navigate safely and appropriately in it.

\clearpage

\setcounter{figure}{0}
\setcounter{table}{0}
\renewcommand{\thefigure}{S\arabic{figure}}
\renewcommand{\thetable}{S\arabic{table}}
\renewcommand{\headeright}{Terwilliger et al.}
\renewcommand{\shorttitle}{Invisible walls --- Supplementary}

\section*{Supplementary}

\subsection*{Methods}

This study and its experimental methods was approved by the University of California, San Diego Institutional Review Board (161452). We obtained informed consent for participants in experiment 4. For Experiments 1, 2, \& 3, we asked for informed consent to be waived. First, we made our pedestrian observations at heavily trafficked locations at a public university campus, where surveillance cameras were already present and visible. As such, these locations do not have the same expectation of privacy as a research laboratory. Second, obtaining consent would have required stopping pedestrians which would have been invasive, disruptive and infeasible.

\subsubsection*{Experiment 1}

We placed two actors on 3 pathways at a university campus (see Fig.\ S1). We used a $2 \times 2 \times 2$ design in which we manipulated whether the actors were looking at each other vs.\ looking at a phone, whether they were talking vs.\ silent, and whether they stood 5 vs.\ 10 feet apart; we also varied actor gender (both men vs.\ both women vs.\ one man and one woman). We filmed a total of 2,319 pedestrians (1,335 men, 921 women, 63 unsure) at different times of day between April 2017 and January 2019. The 3 pathways were next to the campus bookstore, the main student center, and leading to the main campus gym.

Each observation session consisted of 8-minute blocks in which actors were instructed to switch conditions every 2 minutes. The order of conditions was counterbalanced across each session. Between 8-minute blocks, actors took 2--3 minute breaks before proceeding to the next block. A single session typically consisted of 3 blocks.

We manually coded attributes and the behaviors of the observed pedestrians. These included presumed pedestrian gender (man, woman, or unsure), their mode of transit (walking, biking, scootering, skateboarding or other). We also coded whether they were a member of a group, the group's size, and whether they were the leading pedestrian in that group. Finally, we coded whether they breached/passed between our actors (0 = no breach/passed around our actors, 1 = breach/passed between our actors) as well as whether pedestrians were never on course to pass between the actors. For the detailed coding scheme see Table~S1. We excluded 458 pedestrians whose gender was coded `Unsure', or who never traversed near the part of the path with our actors, e.g., if they were always walking on the far outer edge they never needed to deviate. This yielded 1,861 pedestrians (1110 men).

Video coding was performed by 7 undergraduate research assistants (RA). Each RA had to achieve an average accuracy of 70\% across all coded variables as measured against a gold standard annotation set from one of the researchers. Our coders achieved an average accuracy of 89.78 $\pm$ 7.92\%.

We manually coded pedestrian gender, mode of transportation (e.g.\ walking vs.\ scootering), possible distractions (looking at a smartphone or talking to someone), and whether pedestrians appeared to be traveling alone or together.

To measure the effects of involvement signs on the likelihood of breaching proxemic norms (passing between our actors), we fit a generalized linear mixed model (GLMM) with a binomial link function (no breach/around the actors = 0; breach/pass between the actors = 1) and fixed effects for gaze, talk, interpersonal distance, gender of the actors, gender of the pedestrian, and an interaction between gaze and talk. We specified random intercepts for location and for date nested within location, allowing the model to account for variability due to the specific location and day of observation -- for example, at one location on one date, students may have been more inclined to believe they had right of way because they had to be on time for an important exam. The GLMM was fitted using the \textit{lme4} R package \citep{51} (version 1.1-37). All data processing and statistical analysis were performed with the R computing software (version 4.5.1).

Using a likelihood ratio test comparing a generalized linear model (GLM) without the hierarchical structure to our full hierarchical model, we found the hierarchical structure significantly improved model fit ($\chi^2 = 109.05$, df = 2, $P < 0.001$). The GLM was fit using base R and the log likelihood ratio test was performed using base R's \texttt{anova} function.

We used marginal means both for the confidence intervals in Fig.\ 2 and for the odds ratio tests. These marginal means were derived from the GLMM above and computed using the \textit{emmeans} R package \citep{52} (version 1.11.2-8). Estimated marginal means (EMM) allow us to investigate model predictions and to statistically contrast predictors. An EMM is computed by holding selected independent variables constant and using the fitted model to predict outcomes averaged over the remaining variables. For example, with an EMM, we estimated the expected probability and standard deviation of a breach for pedestrians when our actors shared mutual gaze while standing 10 feet apart in silence, averaged over pedestrian gender.

In Fig.\ 2, we plotted estimated marginal means (EMMs) without averaging across all transportation types. Instead, we present model predictions only for pedestrians who were walking, as this reflects the most canonical sense of what a pedestrian is. Including other transportation types does not alter the qualitative pattern of results. Moreover, because the odds ratio--based statistical comparisons are invariant to how non-focal predictors are marginalized over, this choice is irrelevant to the reported effects.

\subsubsection*{Experiment 2}

We placed two women actors, standing 10 feet apart, on a pedestrian pathway at a university campus. We manipulated the actors' mutual body orientation. They stood directly face to face, back to back, or so that they were both 45\textdegree{} offset from the face to face condition (see Fig.\ S2). We also observed pedestrian traffic when no actors were present to obtain baseline traffic patterns.

The research actors were instructed to switch conditions every 10 minutes. The order of conditions was counterbalanced across each session to control for minute-within-the-hour effects resulting from how students move on campus as a consequence of their class schedules. A typical session lasted one hour and included 6 conditions. We filmed a total of 978 pedestrians (458 men) at various times of day between November and December in 2023. Some observed pedestrians did not cross from one side of the space to the other. We included these pedestrians in counts for the crowding predictor but excluded them from our analysis of breaching. This exclusion process yielded 686 pedestrians (301 men).

In order to obtain a wide field of view, each session was filmed using two cameras mounted with mounting tape to a pole. We manually synchronized these videos and produced one side-by-side composited video using FFmpeg. Finally, we used a Python script to draw an anonymized unique identifier above each pedestrian as they moved across the video (see Fig.\ S3).

We manually coded attributes and the behaviors of the observed pedestrians. These included presumed pedestrian gender (man or woman), their mode of transit (walking, biking, electric scootering, etc.), and whether they were distracted (looking at their phone or wearing headphones). We also coded whether they were a member of a group, the group's size, and whether they were the leading pedestrian in that group. Finally, we coded whether they breached/passed between our actors (0 = no breach/passed around our actors, 1 = breach/passed between our actors). For the detailed coding scheme see Table~S2.

Video coding was performed by two coders, JT and JD, who each coded the entire dataset and initially achieved moderate to almost perfect Cohen's Kappa agreement across all variables (mean $\kappa = 0.84$; [0.60, 0.96]). A subsample of disagreements were discussed between JT and JD. A second pass improved Cohen's Kappa agreement across all variables (mean $\kappa = 0.92$; [0.76, 0.99]).

To analyze effects on proxemic norm breaching, we fit a GLMM similar to the procedure in Experiment 1 but with random intercepts for date and minute-within-the-hour. Using a likelihood ratio test comparing a GLM without the hierarchical structure and our full hierarchical model, we found the hierarchical structure significantly improved model fit ($\chi^2 = 6.82$, df = 2, $P = 0.033$).

In Fig.\ 2, we plotted estimated marginal means (EMMs) without averaging across all transportation types. Instead, we present model predictions only for pedestrians who were (a) walking, (b) with no crowd, (c) with no breaches from other pedestrians, and (d) who were not going to or coming from the area behind the camera. This estimate measures the breaching behavior of the prototypical pedestrian. Including other transportation types does not alter the qualitative pattern of results.

\paragraph{Calculating predictors and controls for collective proxemic norm breaching.}

We measured the effects of crowding and behavioral contagion. We measured crowding by counting how many pedestrians were in front of the art mural at the moment a pedestrian passed the actors (excluding the target pedestrian). The square-root of this count was used as a predictor in our GLMM. For behavioral contagion, we measured possible exposures to breaching behavior by counting how many other pedestrians had breached 5 seconds prior to the moment a pedestrian passed the actors. The square-root of this count was used as a predictor in our GLMM. For a pedestrian moving at the average 20-year-old human walking pace of 4.43 feet/second \citep{54}, this would mean behavioral exposures occur 22.15 feet ahead of the average pedestrian. Our results for behavioral contagion do not depend on a 5-second exposure window. They manifest for exposure windows as short as 0.9 seconds and windows up to at least 10 seconds, which is at the limit of plausible direct behavioral exposure.

In Fig.\ S4, we plot the GLMM model's estimated marginal breaching probability as a function of crowding and breaches for the example provided in Fig.\ 5.

\paragraph{Pedestrian trajectory measurement.}

We measured distances between people in continuous world coordinates rather than image-based coordinates (in pixels) or coarse manually labeled categories. To measure this, we built a computer vision pipeline (see Fig.\ S5) which relied on finding a linear mapping (homography) between pixel coordinates in our camera's image and coordinates on a ground plane in the world. With this homography, we could then determine where in the world a pedestrian was located if we knew where their feet were located in an image. The average hold-one-out reprojection error per camera per participant was 15.48\,cm.

In order to measure trajectories, we first detected the location of pedestrians in each video frame as well as the location of their feet using the default pretrained object \& pose estimation weights from YOLOv7 \citep{55}, a fast convolutional neural network architecture. We tracked pedestrians across video frames using Simple Online Realtime Tracking (SORT) \citep{56}, and manually deleted tracking errors using a custom annotation tool. Finally, we imputed missing data, caused by visual obstructions or tracking errors, using a Kalman filter and Rauch--Tung--Striebel smoothing.

\subsubsection*{Experiment 3}

We placed one actor in front of either an informational sign or an art mural and manipulated whether the actor was looking up/out at the sign versus down at their phone (see Fig.\ S6). We also varied the gender of the actor (man vs.\ woman). Experiment 3 was conducted at the same 4 locations as experiments 1 and 2. We observed a total of 1,515 adults (889 men, 593 women, 33 unsure) who passed through 4 locations at UC San Diego at various times of day between August 2018 and April 2019.

Each observation session consisted of 8-minute blocks in which actors were instructed to switch conditions every 2 minutes. The order of conditions was counterbalanced across each session. Between 8-minute blocks, actors took 2--3 minute breaks before proceeding to the next block. A single session typically consisted of 3 blocks.

We manually coded attributes and the behaviors of the observed pedestrians. These included presumed pedestrian gender (man, woman, or unsure), their mode of transit (walking, biking, scootering, skateboarding or other). We also coded whether they were a member of a group, the group's size, and whether they were the leading pedestrian in that group. Finally, we coded whether they breached/passed between our actors (0 = no breach/passed around our actors, 1 = breach/passed between our actors) as well as whether pedestrians were never on course to pass between the actors. For the detailed coding scheme see Table~S1. We excluded 641 pedestrians whose gender was coded `Unsure' ($N = 33$), or who never traversed near the part of the path with our actors ($N = 613$). Note these two categories are not mutually exclusive. This yielded 874 pedestrians (528 men).

Experiment 3 used a similar video coding and statistical procedure as experiment 1. We found the hierarchical structure significantly improved model fit ($\chi^2 = 27.34$, df = 2, $P < 0.001$).

\subsubsection*{Experiment 4}

Participants were recruited from the University of California, San Diego and were compensated with course credit. We tested a total of 99 undergraduate students (46 men, 53 women).

Participants were instructed to walk through an outdoor hallway from a testing room to an outdoor art mural installed on a neighboring building. At the mural they had to wait for 1 minute, and then return to the testing room. Participants were led to believe they were participating in a visual-spatial memory study in which they had to commit the mural to memory in order to sketch it later, all while remembering the route to and from the mural. On their return, we placed either two women actors, two chairs, or nothing (in a baseline condition) in front of the outdoor hallway and observed whether participants passed between or around these obstacles (see Fig.\ S7). We tracked participants' walking trajectories as they approached the outdoor hallway in order to measure how close they approached before changing their trajectory to walk around these obstacles. Trajectories were inferred using the same methodology as in Experiment 2.

In a debriefing interview, we asked participants what they believed the aim of the study was, whether they noticed any cameras, people, or objects along their walk back from the mural and, if they walked between our actors, why they did so.

We fit a Bayesian GLMM model using the \textit{brms} R package \citep{57} (version 2.23.0). Because all participants breached the baseline and chairs conditions, this caused difficulty fitting our model with \textit{lme4}. When fitting this model, we set informative priors on the model intercept and coefficients for obstacle type. This model included random effects for the date of the trial. A Bayes factor analysis revealed that this did not significantly improve model fit, nor did it yield a worse fit (BF = 0.25).

To test whether men were more likely than women pedestrians to expect the actors to yield their interactional space or suspend their interaction, we tested whether men walked closer to the actors than women did before they changed course to walk around the actors. Specifically, we compared how close men and women were to the actors at the moment they changed course to walk around them. For each approaching pedestrian, we linearly extrapolated their imagined path assuming they continued on their current bearing. We then identified points along the trajectory where extrapolated paths switched from breaching to not breaching. We picked the last such point before a pedestrian passed the actors (see Fig.\ 4). Distance to the actors was measured as the distance from the pedestrian at this change-point and the point between the two actors. We fit a linear model using the \texttt{lm} function in the standard library of R. Change-point distance was our dependent variable; pedestrian gender was our independent variable. A log likelihood ratio test found no significant difference between a linear mixed model (LMM) with random intercepts for trial date and a simpler linear model with no random effect structure. However, the Akaike information criterion slightly favored the simpler linear model.

\subsubsection*{Behavioral Coding: Experiments 1 \& 3}

We used the coding scheme in Table~S1 for experiments 1 and 3. For every pedestrian in a recorded video, research assistants coded the following behavioral variables into spreadsheets.

\subsubsection*{Behavioral Coding: Experiment 2}

We used the coding scheme in Table~S2 for experiment 2. Before manual coding, pedestrians were assigned numeric IDs which were overlaid on the videos (see Fig.\ S3). JT and JD coded the following behavioral variables into spreadsheets.

\subsubsection*{Behavioral Coding: Experiment 4}

We used the coding scheme in Table~S3 for experiment 4. JT coded the following behavioral variables into a spreadsheet.

\subsection*{Gender Effects}

We observed equivocal effects of gender on pedestrian navigation. We observed an effect of actor gender in some experiments, but not all. In experiment 1, pedestrians were twice as likely to breach the interactional territory of two women actors than of two men actors (OR = 2.12 $\pm$ 0.61, $z = 2.6$, $P = 0.009$). We did not observe this effect in experiment 3, but did observe it when pooling results from Experiments 1 \& 3 (OR = 1.51 $\pm$ 0.24, $z = 2.59$, $P = 0.010$) (Fig.\ S8).

However, we observed no evidence that pedestrian gender had an effect on breaching in Experiments 1--4, and we observed no interaction between pedestrians' and actors' gender on breaching.

Even if pedestrians do not breach at different rates by gender, it is still possible that gender differences manifest more subtly in walking trajectories. For example, a pedestrian may approach closer before altering their course if they are more likely to expect others will yield for them. To test this, we measured the distance at which pedestrians aimed themselves around our actors in Experiment 4. Pedestrians changed course 28.77 $\pm$ 36.68 feet away from the center of our actors' interactional territory (Fig.\ S9) -- past what Hall calls far public-distance (25 feet) where recognition between acquaintances is common \citep{18}. But, we found no significant difference in the distance at which men and women pedestrians changed course.

\subsection*{The equivocal history of pedestrian gender effects and other social categories}

The absence of pedestrian gender effects is surprising in light of prior work on the proxemics of race \citep{22,23} and gender \citep{23,24,25}. But, we note social category effects are inconsistent between locations within prior studies \citep{22} and between different studies \citep{23,24,25}. Although the inconsistencies in prior work may be partly explained by imprecise \citep{24,25} or unitless spatial measurement \citep{22}, it may be that less privileged social categories may simultaneously result in decreased and increased proxemic spacing. Social categories are flexible resources people may use to enact their sense of social order. One pedestrian may intrude into others' space because of their perceived dominance or power over another, while at the same time, a second pedestrian may avoid intrusions because of `gallantry' \citep{23,58} or a social aversion. Furthermore, these effects likely vary between different cultures, locations, and points in time and even within individuals.

\subsection*{Experiment 3: main effects of interaction type}

Here, we report the main effect of interaction type (actor$\leftrightarrow$actor, actor$\rightarrow$sign, actor$\rightarrow$mural). In the sign condition, pedestrians were 1/4 as likely to breach when the actor gazed up at a sign than when the actor looked down at their phone (OR = 0.22 $\pm$ 0.05, $z = -6.74$, $P < 0.001$). In contrast, in the mural condition, pedestrians were not significantly less likely to breach when the actor gazed up at the mural than when the actor looked down at their phone (OR = 0.86 $\pm$ 0.23, $z = -0.54$, $P = 0.588$).

Because we observed no significant interaction between the art mural condition and directed gaze, these results suggest that pedestrians act as if the proxemic norms imposed by a person viewing an art mural are weaker than those imposed by a social interaction between two people or between a person reading an informational sign. We speculate the difference between the informational sign condition and the mural condition is due to differences in the affordance these media offer readers and viewers. The art mural is large and affords many possible vantage points. At close vantage points, the mural occupies a large visual angle for the viewer; it cannot be simultaneously appreciated in full detail nor can it easily be wholly occluded by any one pedestrian. Whereas, the size of the text on the informational sign afforded many fewer vantage points. A reader would prefer to stand directly in front of it at a close distance. It could easily be wholly occluded by any pedestrian. But recall that we observed an additive interaction between the informational sign condition and directed gaze (OR = 0.47 $\pm$ 0.12, $z = -3.03$, $P = 0.003$). While pedestrians were 3.83 times more likely to breach in the sign condition than in the two actors condition when the actor(s) gazed at their phone (OR = 3.57 $\pm$ 1.20, $z = 3.80$, $P < 0.001$), pedestrians were not significantly more likely to breach in the sign condition than the human condition when actors directed their gaze at a sign/each other (OR = 1.67 $\pm$ 0.56, $z = 1.52$, $P = 0.128$). Together, these results suggest that pedestrians act as if the proxemic norms imposed by a social interaction between two people and those imposed by a person reading an informational sign are similarly strong. However, they treat the proxemic norms imposed by a person standing in front of an informational sign as much weaker.

\begin{figure}[htbp]
  \centering
  \includegraphics[width=\linewidth]{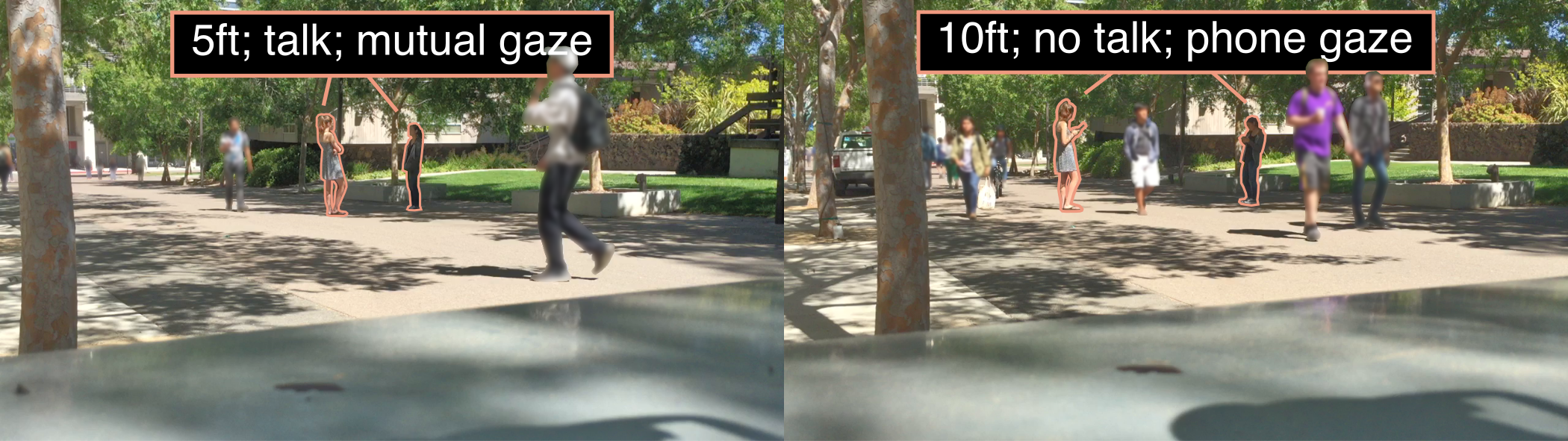}
  \caption{Example video frames showing the experimental manipulations for signs of involvement used in experiment 1.}
  \label{fig:s1}
\end{figure}

\begin{figure}[htbp]
  \centering
  \includegraphics[width=\linewidth]{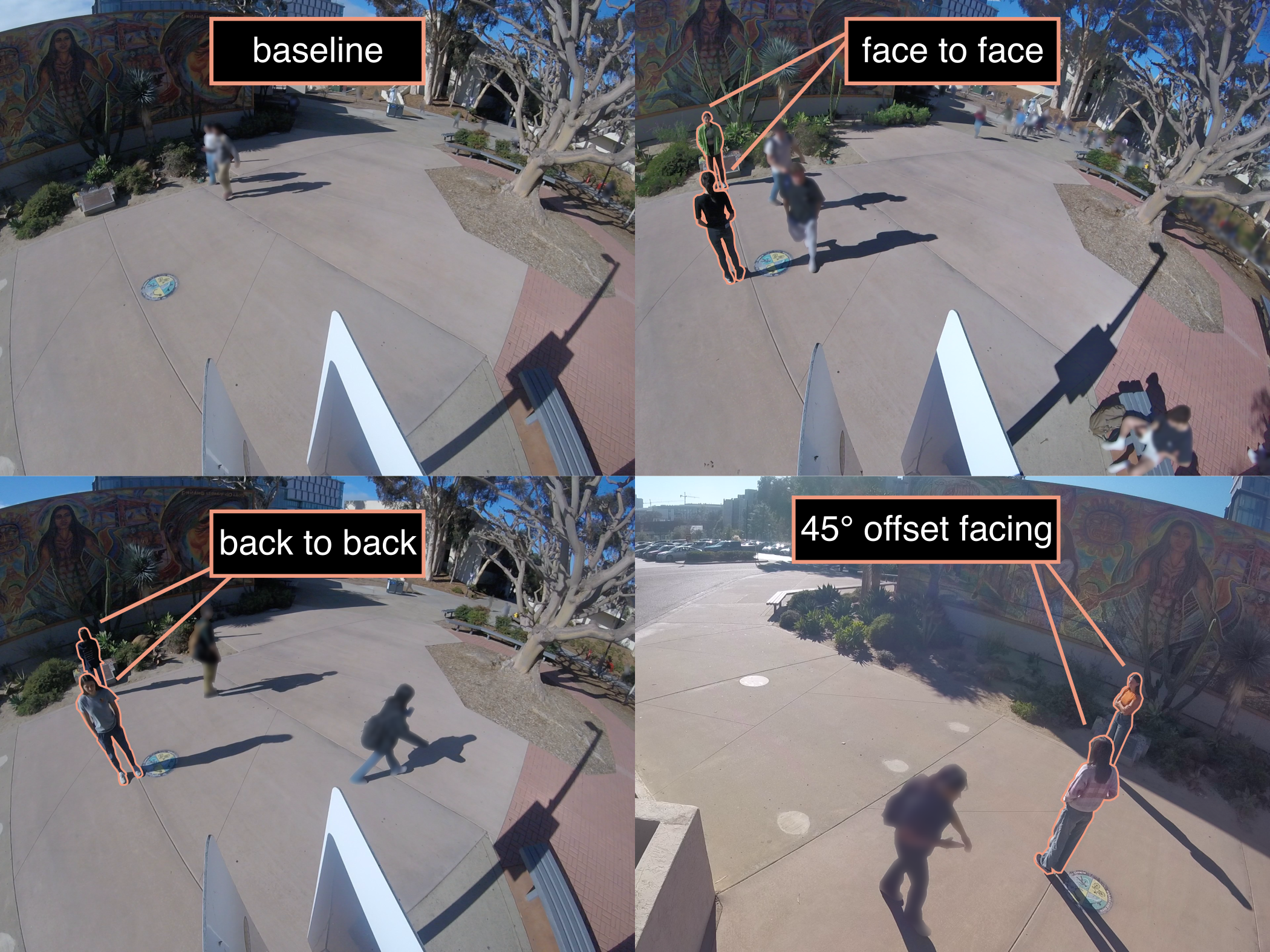}
  \caption{Example video frames showing the experimental manipulations from experiment 2 as well as examples of phone use and non-phone use.}
  \label{fig:s2}
\end{figure}

\begin{figure}[htbp]
  \centering
  \includegraphics[width=\linewidth]{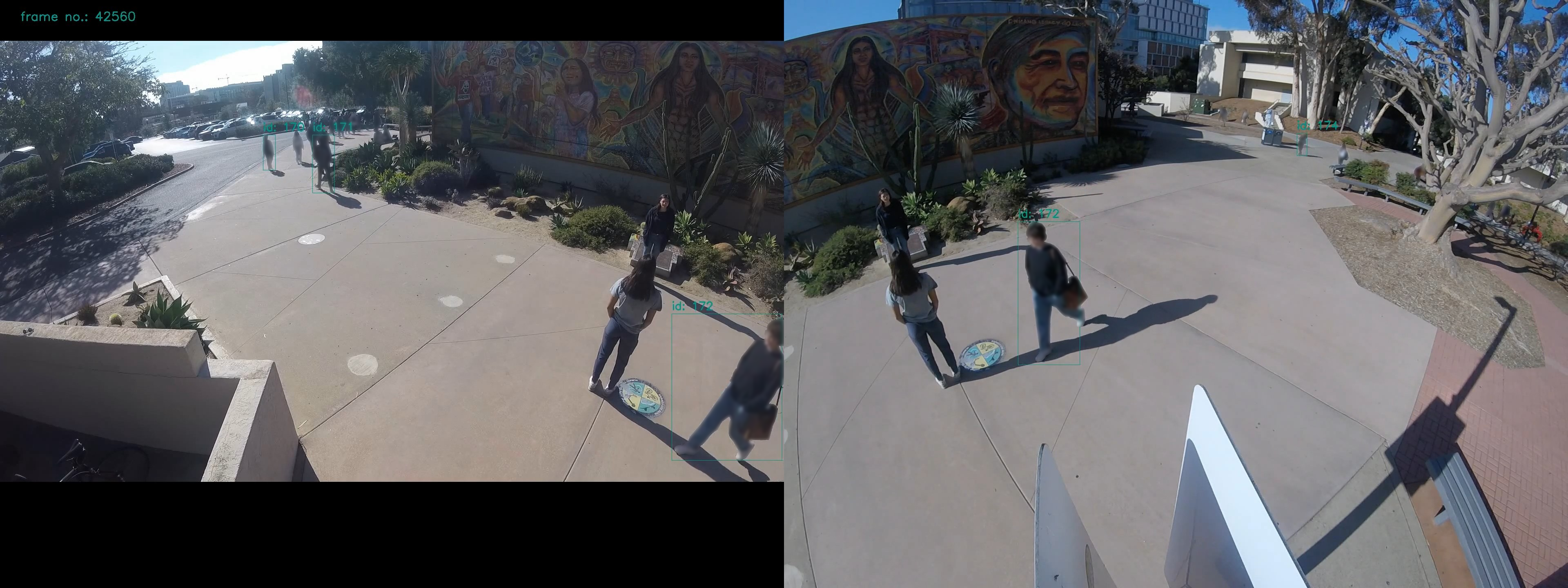}
  \caption{Example video frames showing the synchronized camera views with anonymized IDs for pedestrians.}
  \label{fig:s3}
\end{figure}

\begin{figure}[htbp]
  \centering
  \includegraphics[width=\linewidth]{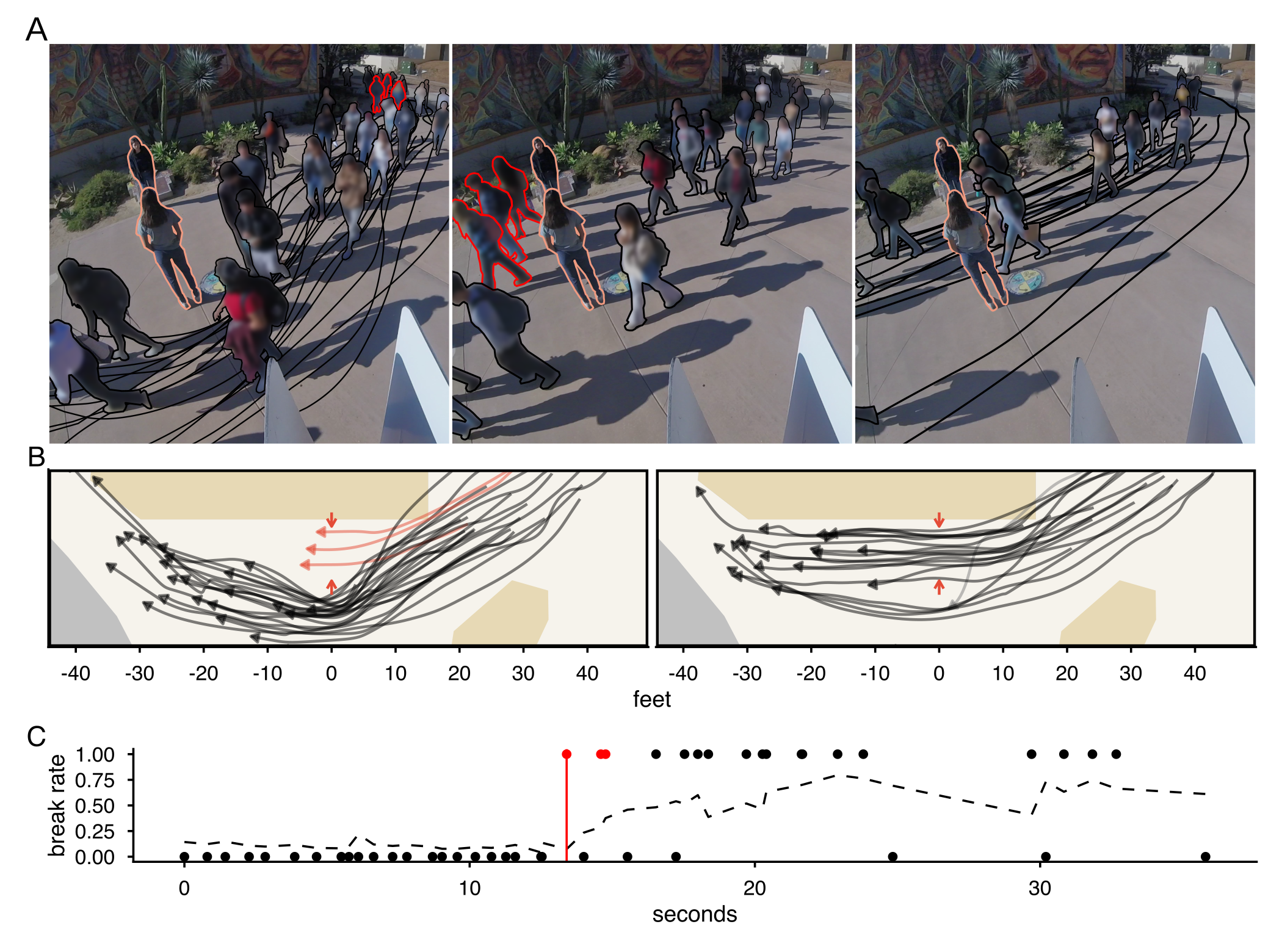}
  \caption{An example of breaching contagion where 46 pedestrians walk past an art mural while two actors silently stood 10 feet apart while facing and staring at each other. Before a group of 3 pedestrians walked between the actors, 22 pedestrians had walked around the actors. After the group of 3 pedestrians walked between the actors, 15 of 19 pedestrians then walked between the actors. (\textbf{A}) Video frames before, during, and after the 3 instigating pedestrians walk between the actors. (\textbf{B}) Trajectories of pedestrians before/during and after the 3 instigating pedestrians walked between the actors. (\textbf{C}) Our statistical model's running predicted breach probability over time which takes into account the number of breaches 5 seconds prior and crowding for a given point in time (black line). Breaches/non-breaches are indicated with circles. The moment the group of three breaches is indicated with a vertical red line.}
  \label{fig:s4}
\end{figure}

\begin{figure}[htbp]
  \centering
  \includegraphics[width=\linewidth]{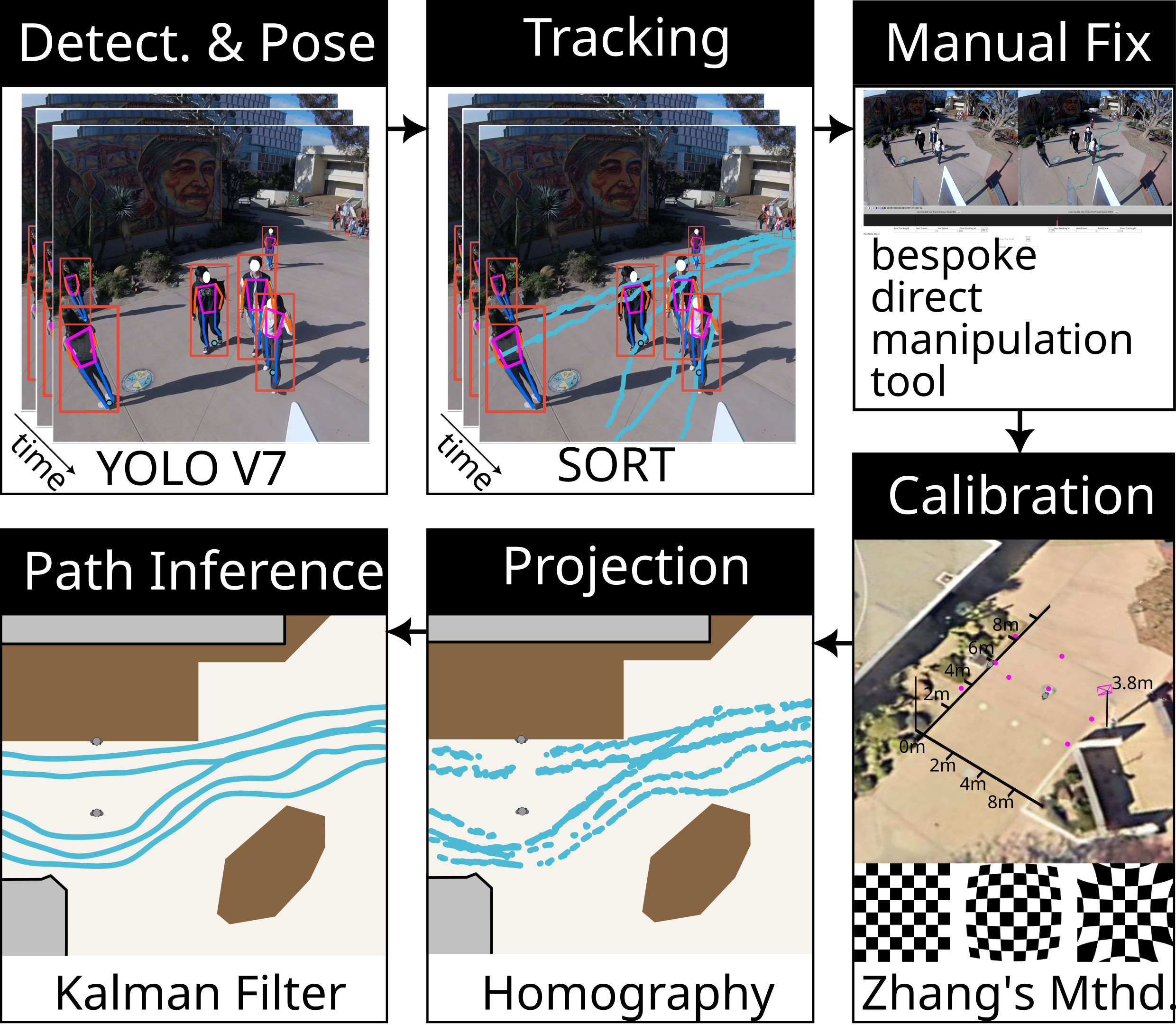}
  \caption{Our trajectory inference method with examples from experiment 2. (1) Results of person detection \& pose estimation are shown. (2) Results of pedestrian tracking are shown with raw pixel-coordinate trajectories shown in blue. (3) A screenshot of our manual annotation tools. (4) To calibrate our camera, we needed to know its intrinsic properties (how the lens distorts light) and where the camera was located in space. Landmarks used in study 4 are shown in magenta. (5) Raw world-coordinate trajectory data is shown in blue. (6) Smoothed and interpolated data is shown in blue.}
  \label{fig:s5}
\end{figure}

\begin{figure}[htbp]
  \centering
  \includegraphics[width=\linewidth]{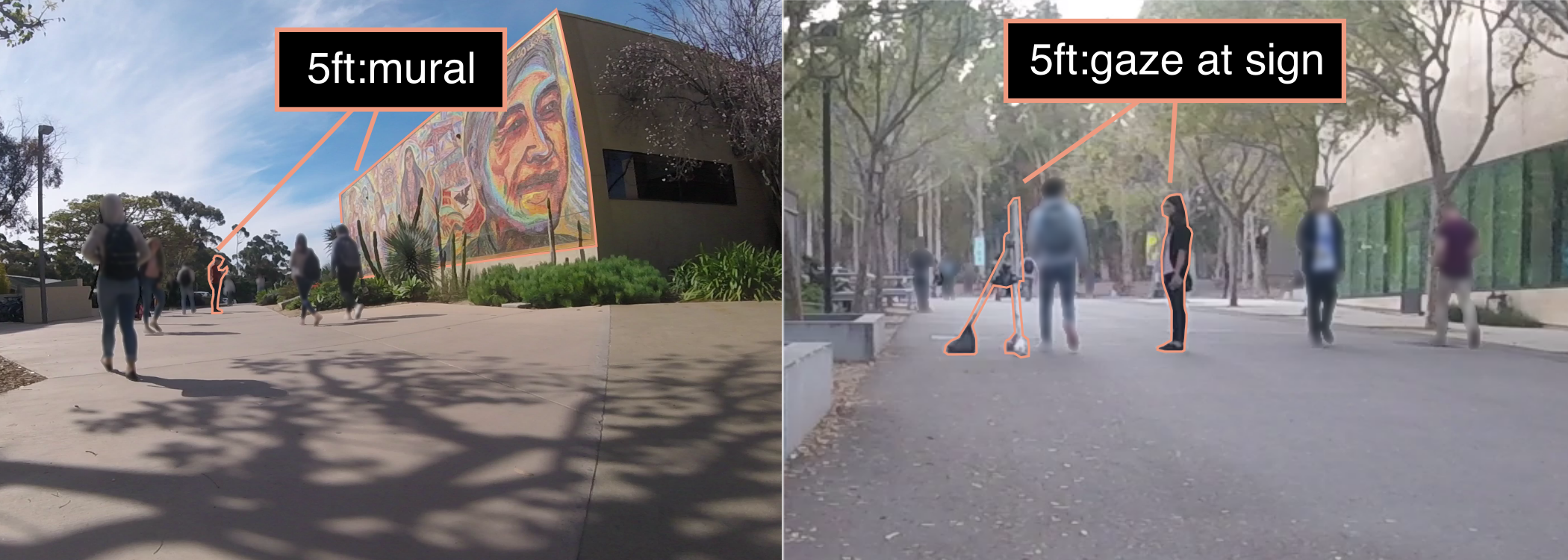}
  \caption{Example video frames showing the experimental manipulations from experiment 3.}
  \label{fig:s6}
\end{figure}

\begin{figure}[htbp]
  \centering
  \includegraphics[width=\linewidth]{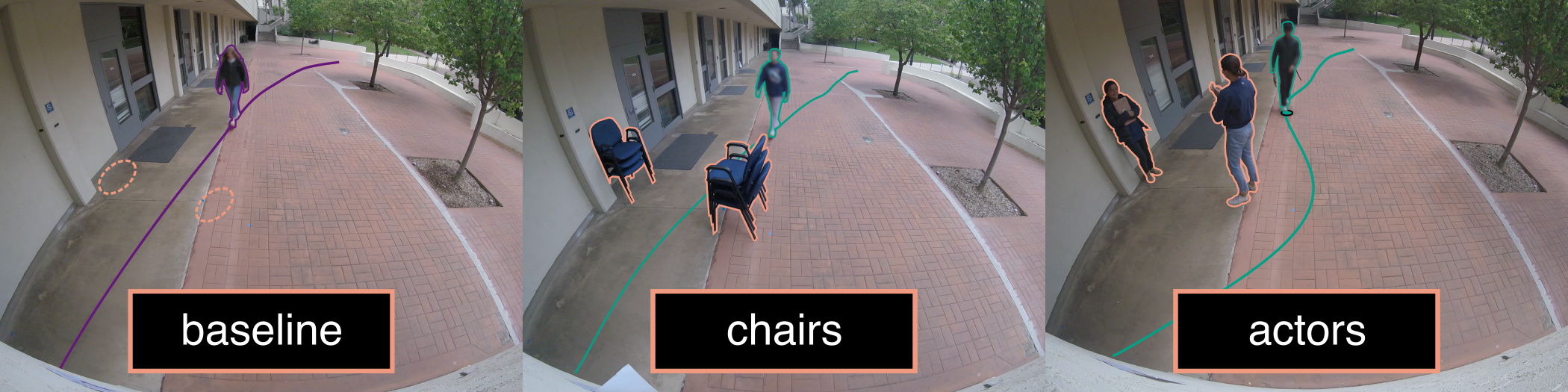}
  \caption{Example video frames showing the experimental manipulations from experiment 4.}
  \label{fig:s7}
\end{figure}

\begin{figure}[htbp]
  \centering
  \includegraphics[width=\linewidth]{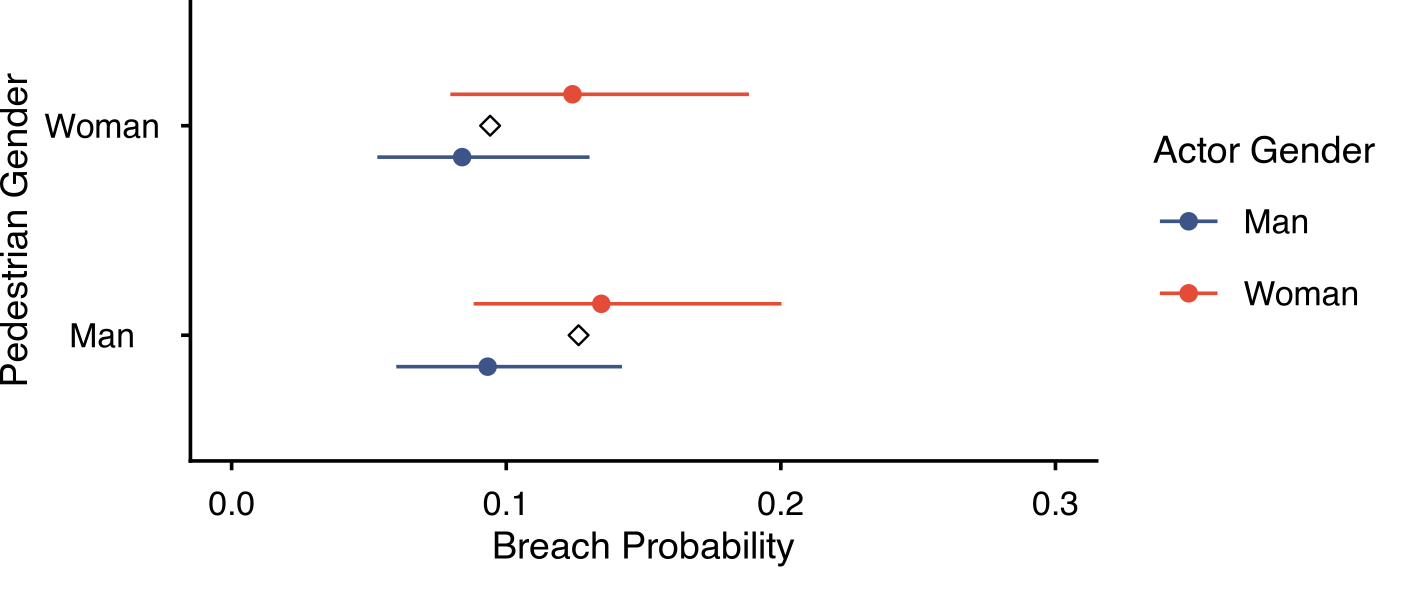}
  \caption{Pooled breach probabilities from Experiment 1 \& 3, when pedestrians stood at 5 feet. Colored dots and lines represent the estimated marginal effects and 95\% confidence intervals for each condition. Empirical rates are shown as white diamonds.}
  \label{fig:s8}
\end{figure}

\begin{figure}[htbp]
  \centering
  \includegraphics[width=\linewidth]{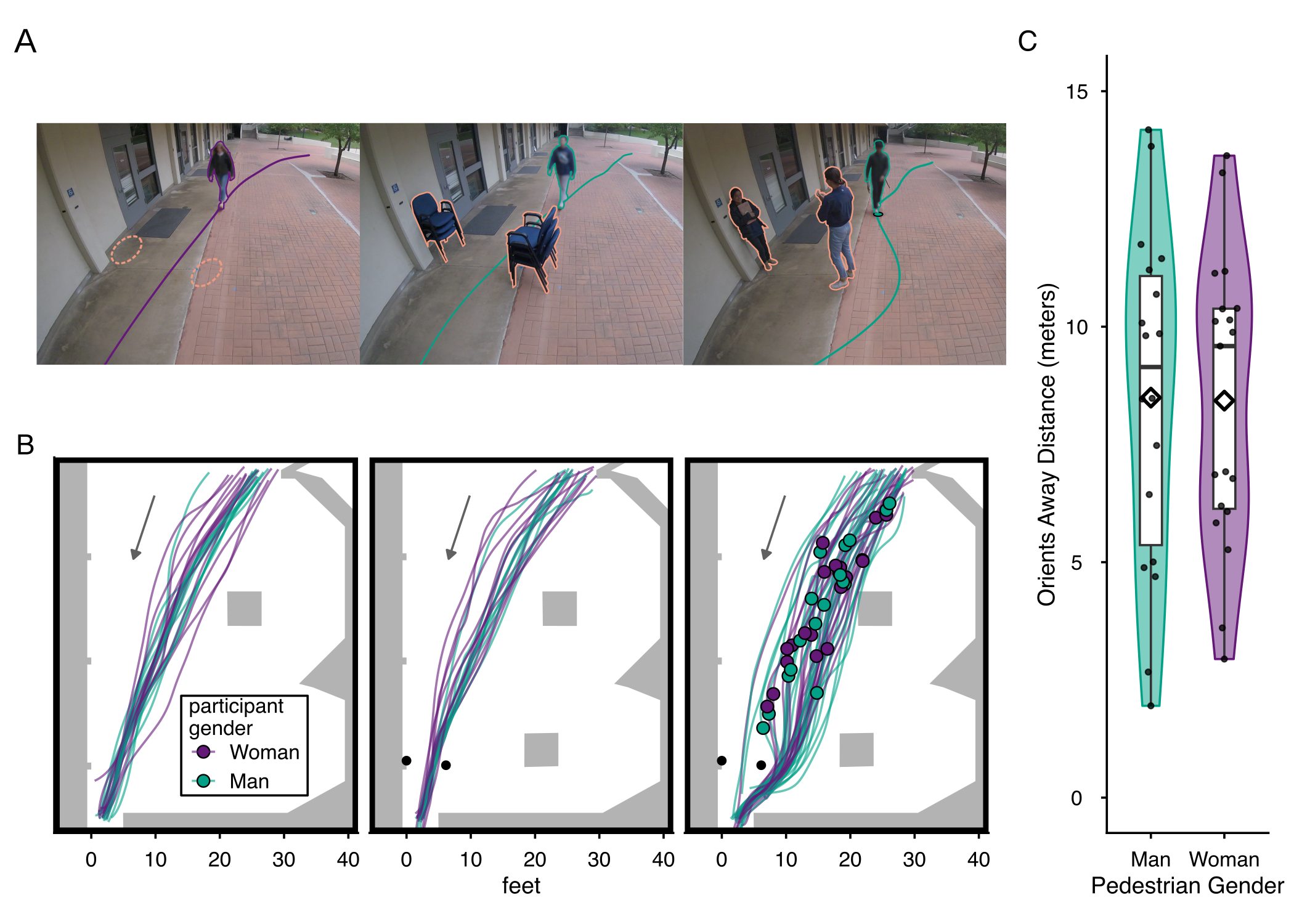}
  \caption{Pedestrian trajectories from baseline, chair, and actor conditions in Experiment 4. Trajectories are colored by pedestrian gender (green for men and purple for women). Circles mark the point at which each pedestrian aimed themselves around the actors.}
  \label{fig:s9}
\end{figure}

\subsection*{Tables}

\begin{table}[htbp]
\caption{The Coding Scheme used in Experiments 1 and 3.}
\label{tab:s1}
\centering
\small
\setlength{\tabcolsep}{5pt}
\renewcommand{\arraystretch}{1.35}
\begin{tabular}{>{\raggedright\arraybackslash}p{2.6cm} >{\raggedright\arraybackslash}p{2.7cm} >{\raggedright\arraybackslash}p{4.7cm} >{\raggedright\arraybackslash}p{3.5cm}}
\toprule
\textbf{Category} & \textbf{Variable} & \textbf{Definition} & \textbf{Options} \\
\midrule
\multirow{7}{=}{Condition Information}
  & Date & The date that the recording took place & month/day/year \\
  & Gaze & Exp 1: are the actors sharing mutual gaze? Exp 3: is the actor looking at the informational panel or mural? & 0 -- no, looking down at phone; 1 -- yes \\
  & Talk & Are the actors talking? & 0 -- no; 1 -- yes \\
  & Actor Gender & The gender of the actor forming the invisible wall & M -- man; W -- woman \\
  & Distance & Distance between the actors & 5 feet; 10 feet \\
  & Location & The designated on-campus location & RIMAC; PC; Bookstore \\
  & Type of interaction & What type of interaction is the actor engaged in? & Human; Sign; Mural \\
\midrule
\multirow{3}{=}{Pedestrian Group Info}
  & Group size & Number of pedestrians within 0.6\,m radius of each other & number \\
  & Group Membership & Group membership ID & numeric/alphabetic code \\
  & Presumed gender & The apparent gender of the pedestrian & M -- man; W -- woman; U -- unsure \\
\midrule
\multirow{5}{=}{Interaction Information}
  & Start (ENTER) & Time pedestrian begins diverging & HH:MM:SS \\
  & End (EXIT) & Time pedestrian crosses $\sim$0.3\,m past the wall & HH:MM:SS \\
  & Outcome & Behavior of the pedestrian & Divergence; Breach; Within bounds, no divergence \\
  & Preoccupation & Potential distraction & Talking on phone; Looking at phone; Headphones; Talking with group; Other; Not preoccupied \\
  & Transportation & Mode of transportation & Walking; Biking; Skateboarding; Scootering; Other \\
\bottomrule
\end{tabular}
\end{table}

\begin{table}[htbp]
\caption{The Coding Scheme used in Experiment 2.}
\label{tab:s2}
\centering
\small
\setlength{\tabcolsep}{5pt}
\renewcommand{\arraystretch}{1.35}
\begin{tabular}{>{\raggedright\arraybackslash}p{2.6cm} >{\raggedright\arraybackslash}p{2.7cm} >{\raggedright\arraybackslash}p{4.7cm} >{\raggedright\arraybackslash}p{3.5cm}}
\toprule
\textbf{Category} & \textbf{Variable} & \textbf{Definition} & \textbf{Options} \\
\midrule
\multirow{3}{=}{Condition Information}
  & Date & The date that the recording took place & month/day/year \\
  & Time & The time that the recording started & HH:MM:SS.fff \\
  & Condition & The scenario the actors were placed in & Baseline; FaceToFace; FacingOffset; BackToBack \\
\midrule
\multirow{3}{=}{Trajectory Information}
  & Passes target region & Does the pedestrian cross the line between the actors? & 0 -- yes; 1 -- no \\
  & Outcome & How does the pedestrian navigate past the actors? & 0 -- does not pass between; 1 -- passes between \\
  & Pedestrian Direction & Direction of pedestrian travel & Left; Right \\
\midrule
\multirow{6}{=}{Pedestrian Information}
  & Person ID & Preassigned unique identifier & Integer \\
  & Pedestrian Gender & Presumed binary gender & M -- man; W -- woman; U -- unsure \\
  & Transportation & Mode of transportation & Walking; Bike; Skateboard; Scooter; Other \\
  & Motorized Transportation & Is transportation motorized? & 0 -- no; 1 -- yes \\
  & Phone use & Is the pedestrian using a phone? & 0 -- no; 1 -- yes \\
  & Headphones & Is the pedestrian wearing headphones? & 0 -- no; 1 -- yes \\
\midrule
\multirow{5}{=}{Pedestrian Group Info}
  & Group ID & Social group identifier & Integer; NA \\
  & Group size & Number of people in the group & Number \\
  & Group lead & First in group to cross the actor line? & 0 -- no; 1 -- yes \\
  & Is inner pedestrian & Closest to the mural? & 0 -- no; 1 -- yes \\
  & Is outer pedestrian & Closest to the camera? & 0 -- no; 1 -- yes \\
\bottomrule
\end{tabular}
\end{table}

\begin{table}[htbp]
\caption{The Coding Scheme used in Experiment 4.}
\label{tab:s3}
\centering
\small
\setlength{\tabcolsep}{5pt}
\renewcommand{\arraystretch}{1.35}
\begin{tabular}{>{\raggedright\arraybackslash}p{2.6cm} >{\raggedright\arraybackslash}p{2.7cm} >{\raggedright\arraybackslash}p{4.7cm} >{\raggedright\arraybackslash}p{3.5cm}}
\toprule
\textbf{Category} & \textbf{Variable} & \textbf{Definition} & \textbf{Options} \\
\midrule
Pedestrian Information & Person ID & Preassigned unique identifier & Integer \\
\midrule
Trajectory Information & Outcome & How does the pedestrian navigate past the actors? & 0 -- does not pass between actors; 1 -- passes between actors \\
\bottomrule
\end{tabular}
\end{table}


\clearpage
\begin{thebibliography}{99}

\bibitem[Tong \& Bode(2022)]{1} Y. Tong, N. W. Bode, The principles of pedestrian route choice. \textit{Journal of the Royal Society Interface} \textbf{19}, (2022). doi:10.1098/rsif.2022.0061

\bibitem[Bacik et~al.(2023)]{2} K. A. Bacik, B. S. Bacik, T. Rogers, Lane nucleation in complex active flows. \textit{Science} \textbf{379}, 923--928 (2023). doi:10.1126/science.add8091

\bibitem[Gu et~al.(2025)]{3} F. Gu, B. Guiselin, N. Bain, I. Zuriguel, D. Bartolo, Emergence of collective oscillations in massive human crowds. \textit{Nature} \textbf{638}, 112--119 (2025). doi:10.1038/s41586-024-08514-6

\bibitem[Helbing \& Moln\'{a}r(1995)]{4} D. Helbing, P. Moln\'{a}r, Social force model for pedestrian dynamics. \textit{Physics Review E} \textbf{51}, 4282--4286 (1995). doi:10.1103/PhysRevE.51.4282

\bibitem[Brown et~al.(2021)]{5} G. L. Brown, N. Seethapathi, M. Srinivasan, A unified energy-optimality criterion predicts human navigation paths and speeds. \textit{Proceedings of the National Academy of Sciences} \textbf{118}, e2020327118 (2021). doi:10.1073/pnas.2020327118

\bibitem[Ma et~al.(2021)]{6} Y. Ma, E. W. M. Lee, M. Shi, R. K. K. Yuen, Spontaneous synchronization of motion in pedestrian crowds of different densities. \textit{Nature Human Behaviour} \textbf{5}, 447--457 (2021). doi:10.1038/s41562-020-00997-3

\bibitem[Gilbert(1990)]{7} M. Gilbert, Walking together: A paradigmatic social phenomenon. \textit{MidWest Studies in Philosophy} \textbf{15}, 1--14 (1990). doi:10.1111/j.1475-4975.1990.tb00202.x

\bibitem[Ryave \& Schenkein(1974)]{8} A. L. Ryave, J. N. Schenkein, ``Notes on the art of walking'' in \textit{Ethnomethodology: Selected Readings} (Penguin, 1974), pp.\ 265--274.

\bibitem[Alexander(2002)]{9} R. M. Alexander, Energetics and optimization of human walking and running: the 2000 Raymond Pearl memorial lecture. \textit{American Journal of Human Biology} \textbf{14}, 641--648 (2002). doi:10.1002/ajhb.10067

\bibitem[Murakami et~al.(2021)]{10} H. Murakami, C. Feliciani, Y. Nishiyama, K. Nishinari, Mutual anticipation can contribute to self-organization in human crowds. \textit{Science Advances} \textbf{7}, eabe7758 (2021). doi:10.1126/sciadv.abe7758

\bibitem[Karamouzas et~al.(2014)]{11} I. Karamouzas, B. Skinner, S. J. Guy, Universal Power Law Governing Pedestrian Interactions. \textit{Physics Review Letters} \textbf{113}, 238701 (2014). doi:10.1103/PhysRevLett.113.238701

\bibitem[Ma et~al.(2025)]{12} Y. Ma, M. Shi, W. Xie, Z. Hu, Y. Wei, T. Zeng, E. W. M. Lee, Unraveling human crowd dynamics through the foot tracking of pedestrians. \textit{Science Advances} \textbf{11}, eadw2688 (2025). doi:10.1126/sciadv.adw2688

\bibitem[Sherry et~al.(1992)]{13} D. F. Sherry, L. F. Jacobs, S. J. Gaulin, Spatial memory and adaptive specialization of the hippocampus. \textit{Trends in Neuroscience} \textbf{15}(8), 298--303 (1992). doi:10.1016/0166-2236(92)90080-R

\bibitem[Nathan et~al.(2008)]{14} R. Nathan, W. M. Getz, E. Revilla, M. Holyoak, R. Kadmon, D. Saltz, P. E. Smouse, A movement ecology paradigm for unifying organismal movement research. \textit{Proceedings of the National Academy of Sciences} \textbf{105}(49), 19052--19059 (2008). doi:10.1073/pnas.0800375105

\bibitem[Strandburg-Peshkin et~al.(2015)]{15} A. Strandburg-Peshkin, D. R. Farine, I. D. Couzin, M. C. Crofoot, Shared decision-making drives collective movement in wild baboons. \textit{Science} \textbf{348}, 1358--1361 (2015). doi:10.1126/science.aaa5099

\bibitem[Sayin et~al.(2025)]{16} S. Sayin, E. Cousin-Fuchs, I. Petelski, Y. G\"{u}nzel, M. Salahshour, C. Y. Lee, L. Li, O. Deussen, G. A. Sword, I. D. Couzin, The behavioral mechanisms governing collective motion in swarming locusts. \textit{Science} \textbf{387}, 995--1000 (2025).

\bibitem[Kaufhold et~al.(2025)]{17} S. P. Kaufhold, J. Terwilliger, F. Rossano, Socially situated navigation: social rank and sex influence spatial navigation strategies in Japanese macaques. \textit{Proceedings of the Annual Meeting of the Cognitive Science Society} vol.\ 47 (2025).

\bibitem[Hall(1966)]{18} E. T. Hall, \textit{The Hidden Dimension} (Doubleday \& Co, 1966).

\bibitem[Sommer(1959)]{19} R. Sommer, Studies in personal space. \textit{Sociometry} \textbf{22}, 247--260 (1959). doi:10.2307/2785668

\bibitem[Zhou et~al.(2022)]{20} C. Zhou, M. C. Miao, X. R. Chen, Y. F. Hu, Q. Chang, M. Y. Yan, S. G. Kuai, Human-behaviour-based social locomotion model improves the humanization of social robots. \textit{Nature Machine Intelligence} \textbf{4}, 1040--1052 (2022). doi:10.1038/s42256-022-00542-z

\bibitem[Sommer(1969)]{21} R. Sommer, \textit{Personal Space. The Behavioral Basis of Design} (Prentice-Hall, 1969).

\bibitem[Dietrich \& Sands(2023)]{22} B. J. Dietrich, M. L. Sands, Seeing racial avoidance on New York City streets. \textit{Nature Human Behaviour} \textbf{7}, 1275--1281 (2023). doi:10.1038/s41562-023-01589-7

\bibitem[Willis et~al.(1979)]{23} F. N. Willis Jr., J. A. Gier, D. E. Smith, Stepping Aside: Correlates of Displacement in Pedestrians. \textit{Journal of Communication} \textbf{29}, 34--39 (1979). doi:10.1111/j.1460-2466.1979.tb01739.x

\bibitem[Dabbs \& Stokes(1975)]{24} J. M. Dabbs, N. A. Stokes, Beauty is Power: The Use of Space on the Sidewalk. \textit{Sociometry} \textbf{38}, 551--557 (1975). doi:10.2307/2786367

\bibitem[Sobel \& Lillith(1975)]{25} R. S. Sobel, N. Lillith, Determinants of Nonstationary Personal Space Invasion. \textit{Journal of Social Psychology} \textbf{97}, 39--45 (1975). doi:10.1080/00224545.1975.9923310

\bibitem[Kendon(1990a)]{26} A. Kendon, ``The F-formation system'' in \textit{Conducting Interaction: Patterns of Behavior in Focused Encounters} (Cambridge University Press, 1990), pp.\ 209--237.

\bibitem[Lyman \& Scott(1967)]{27} S. M. Lyman, M. B. Scott, Territoriality: A Neglected Sociological Dimension. \textit{Social Problems} \textbf{15}, 236--249 (1967). doi:10.1525/sp.1967.15.2.03a00090

\bibitem[Goffman(1963)]{28} E. Goffman, \textit{Behavior in Public Spaces} (The Free Press, 1963).

\bibitem[Levinson(2025)]{29} S. C. Levinson, ``The Interaction Engine and Social Life'' in \textit{The Interaction Engine: Language in Social Life and Human Evolution} (Cambridge University Press, 2025), pp.\ 107--148. doi:10.1017/9781009570343.006

\bibitem[Dingemanse et~al.(2023)]{30} M. Dingemanse \textit{et al.}, Beyond Single-Mindedness: A Figure-Ground Reversal for the Cognitive Sciences. \textit{Cogn.\ Sci.} \textbf{47}, e13230 (2023). doi:10.1111/cogs.13230

\bibitem[Zhou et~al.(2019)]{31} C. Zhou, M. Han, Q. Liang, Y. F. Hu, S. G. Kuai, A social interaction field model accurately identifies static and dynamic social groupings. \textit{Nature Human Behaviour} \textbf{3}, 847--855 (2019). doi:10.1038/s41562-019-0618-2

\bibitem[Setti et~al.(2015)]{32} F. Setti, C. Russell, C. Bassetti, M. Cristani, F-Formation Detection: Individuating Free-Standing Conversational Groups in Images. \textit{PLOS One} \textbf{10}, e0123783 (2015). doi:10.1371/journal.pone.0123783

\bibitem[McMahon \& Isik(2023)]{33} E. McMahon, L. Isik, Seeing social interactions. \textit{Trends in Cognitive Science} \textbf{27}, 1165--1179 (2023). doi:10.1016/j.tics.2023.09.001

\bibitem[Skripkauskaite et~al.(2023)]{34} S. Skripkauskaite, I. Mihai, K. Koldewyn, Attentional bias towards social interactions during viewing of naturalistic scenes. \textit{Quarterly Journal of Experimental Psychology} \textbf{76}, 2303--2311 (2023). doi:10.1177/17470218221140879

\bibitem[Vestner et~al.(2019)]{35} T. Vestner, S. P. Tipper, T. Hartley, H. Over, S. A. Rueschemeyer, Bound together: Social binding leads to faster processing, spatial distortion, and enhanced memory of interacting partners. \textit{Journal of Experimental Psychology: General} \textbf{148}, 1251--1268 (2019). doi:10.1037/xge0000545

\bibitem[Ding et~al.(2017)]{36} X. Ding, Z. Gao, M. Shen, Two Equals One: Two Human Actions During Social Interaction Are Grouped as One Unit in Working Memory. \textit{Psychological Science} \textbf{28}, 1311--1320 (2017). doi:10.1177/0956797617707318

\bibitem[Brown \& Levinson(1987)]{37} P. Brown, S. C. Levinson, \textit{Politeness: Some Universals in Language Usage} (Cambridge University Press, 1987).

\bibitem[Goffman(1955)]{38} E. Goffman, On face-work: an analysis of ritual elements in social interaction. \textit{Psychiatry} \textbf{18}, 213--231 (1955).

\bibitem[Krause et~al.(2021)]{39} J. Krause, P. Romanczuk, E. Cracco, W. Arlidge, A. Nassauer, M. Brass, Collective rule-breaking. \textit{Trends in Cognitive Sciences} \textbf{25}, 1082--1095 (2021). doi:10.1016/j.tics.2021.08.003

\bibitem[Faria et~al.(2010)]{40} J. J. Faria, S. Krause, J. Krause, Collective behavior in road crossing pedestrians: the role of social information. \textit{Behavioral Ecology} \textbf{21}, 1236--1242 (2010). doi:10.1093/beheco/arq141

\bibitem[Keizer et~al.(2008)]{41} K. Keizer, S. Lindenberg, L. Steg, The Spreading of Disorder. \textit{Science} \textbf{322}, 1681--1685 (2008). doi:10.1126/science.1161405

\bibitem[Garfinkel(1967)]{42} H. Garfinkel, \textit{Studies in Ethnomethodology} (Prentice-Hall, 1967).

\bibitem[Rossano(2012)]{43} F. Rossano, ``Gaze in Conversation'' in \textit{The Handbook of Conversation Analysis} J. Sidnell, T. Stivers, Eds. (Wiley, 2012) pp.\ 308--329. doi:10.1002/9781118325001.ch15

\bibitem[Anderson(1981)]{44} N. H. Anderson, \textit{Foundations of Information Integration Theory} (Academic Press, 1981).

\bibitem[Kendon(1990b)]{45} A. Kendon, ``Description of Some Human Greetings'' in \textit{Conducting Interaction: Patterns of Behavior in Focused Encounters} (Cambridge University Press, 1990), pp.\ 153--208.

\bibitem[Papeo et~al.(2019)]{46} L. Papeo, N. Goupil, S. Soto-Faraco, Visual Search for People Among People. \textit{Psychological Science} \textbf{30}, 1483--1496 (2019). doi:10.1177/0956797619867295

\bibitem[Emery(2000)]{47} N. J. Emery, The eyes have it: the neuroethology, function and evolution of social gaze. \textit{Neuroscience \& Biobehavioral Reviews} \textbf{24}, 581--604 (2000). doi:10.1016/S0149-7634(00)00025-7

\bibitem[{von Uexk\"{u}ll}(1957)]{48} J. von Uexk\"{u}ll, ``A stroll through the worlds of animals and men: A picture book of invisible worlds'' in \textit{Instinctive Behavior: The Development of a Modern Concept} C. H. Schiller, Ed. and Transl. (International Universities Press, 1957), pp.\ 5--80.

\bibitem[Rios-Martinez et~al.(2011)]{49} J. Rios-Martinez, A. Spalanzani, C. Laugier, ``Understanding human interaction for probabilistic autonomous navigation using Risk-RRT approach'' in \textit{2011 IEEE/RSJ International Conference on Intelligent Robots and Systems} (2011), pp.\ 2014--2019. doi:10.1109/IROS.2011.6094496

\bibitem[Ettinger et~al.(2021)]{50} S. Ettinger \textit{et al.}, Large Scale Interactive Motion Forecasting for Autonomous Driving: The Waymo Open Motion Dataset. \textit{arXiv}:2104.10133 [cs] (2021). doi:10.48550/arXiv.2104.10133

\bibitem[Bates et~al.(2015)]{51} D. Bates, M. M\"{a}chler, B. Bolker, S. Walker, Fitting Linear Mixed-Effects Models Using lme4. \textit{Journal of Statistical Software} \textbf{67}, 1--48 (2015). doi:10.18637/jss.v067.i01

\bibitem[Lenth(2025)]{52} R. Lenth, \textit{emmeans}: Estimated marginal means, aka least-squares means. R package version 1.11.2-8 (2025).

\bibitem[Bohannon \& Williams(2011)]{54} R. W. Bohannon, A. Williams, Normal walking speed: a descriptive meta-analysis. \textit{Physiotherapy} \textbf{97}, 182--189 (2011). doi:10.1016/j.physio.2010.12.004

\bibitem[Wang et~al.(2023)]{55} C. Y. Wang, A. Bochkovskiy, H. Y. M. Liao, ``YOLOv7: Trainable Bag-of-Freebies Sets New State-of-the-Art for Real-Time Object Detectors'' in \textit{IEEE/CVF Conference on Computer Vision and Pattern Recognition} (2023), pp.\ 7464--7475. doi:10.1109/CVPR52729.2023.00721

\bibitem[Bewley et~al.(2016)]{56} A. Bewley, Z. Ge, L. Ott, F. Ramos, B. Upcroft, ``Simple online and realtime tracking'' in \textit{2016 IEEE International Conference on Image Processing (ICIP)} (2016), pp.\ 3464--3468. doi:10.1109/ICIP.2016.7533003

\bibitem[B\"{u}rkner(2017)]{57} P. C. B\"{u}rkner, brms: An R Package for Bayesian Multilevel Models Using Stan. \textit{Journal of Statistical Software} \textbf{80}, 1--28 (2017). doi:10.18637/jss.v080.i01

\bibitem[Goffman(2009)]{58} E. Goffman, \textit{Relations in Public} (Transaction Publishers, 2009).

\end{thebibliography}
\end{document}